\documentclass{article}

\usepackage{lineno}
\usepackage{ae,aecompl}
\usepackage[utf8]{inputenc}
\usepackage[T1]{fontenc}
\usepackage[english]{babel}
\usepackage[style=numeric-comp,backend=biber,natbib=true,sorting=none]{biblatex}
\usepackage{csquotes}
\usepackage{amssymb, amsmath, amsthm, amsfonts}
\usepackage{mathtools}
\usepackage{siunitx}
\usepackage{caption}
\usepackage{multirow}
\usepackage{derivative}
\usepackage{physics}
\usepackage[hidelinks]{hyperref}
\usepackage{enumerate, cleveref}
\usepackage{tabstackengine}
\usepackage{stackengine}
\usepackage{algorithm, algpseudocode}
\usepackage{supertabular, makecell, booktabs,wrapfig, longtable}
\usepackage{orcidlink}
\usepackage{textcomp}
\usepackage{subcaption}
\usepackage{placeins, relsize, fix-cm}
\usepackage{todonotes}
\usepackage{tikz}
\usetikzlibrary{shapes.geometric, arrows, patterns, patterns.meta, calc}

\usepackage{geometry}
\usepackage{authblk}
\usepackage{aliascnt}

\numberwithin{equation}{section}

\newtheoremstyle{remarkbreak}%
  {6pt}   
  {6pt}   
  {\itshape}  
  {6pt}      
  {\bfseries} 
  {}      
  {\newline} 
  {%
    \thmname{#1}\thmnumber{ #2}%
    \if\relax\detokenize{#3}\relax\else: #3\fi
    \hspace*{\parindent} 
}

\theoremstyle{remarkbreak}

\newaliascnt{remark}{theorem}

\aliascntresetthe{remark}
\newtheorem*{remark*}{Remark}

\crefname{remark}{remark}{remarks}
\Crefname{remark}{Remark}{Remarks}

\numberwithin{equation}{section}

\patchcmd{\subequations}  
{\theparentequation\alph{equation}}  
{\theparentequation.\alph{equation}}  
{}{}  
\DeclareUnicodeCharacter{2212}{\textminus}

\newcommand{\numof}[1]{N_{#1}}
\newcommand{\primvars}{X}  
\newcommand{\secvars}{Y}  
\newcommand{\cidx}{\zeta}  
\newcommand{\cridx}{\eta}  
\newcommand{\pidx}{\beta}  
\newcommand{\pridx}{\beta}  

\newcommand{\ofracSymb}{z}
\newcommand{\ofrac}[1][\cidx]{\ofracSymb_{\if\relax\detokenize{#1}\relax\else#1\fi}}
\newcommand{\pfrac}[1][\pidx]{y_{\if\relax\detokenize{#1}\relax\else#1\fi}}
\newcommand{\sat}[1][\pidx]{s_{\if\relax\detokenize{#1}\relax\else#1\fi}}
\newcommand{\cpfrac}[1][\cidx\pidx]{x_{\if\relax\detokenize{#1}\relax\else#1\fi}}
\newcommand{\ecpfrac}[1][\cidx\pidx]{\chi_{\if\relax\detokenize{#1}\relax\else#1\fi}}
\newcommand{\chempot}[1][\cidx\pidx]{\mu_{\if\relax\detokenize{#1}\relax\else#1\fi}}
\newcommand{\fugcoeffSymb}{\varphi}
\newcommand{\fugcoeff}[1][\cidx\pidx]{\fugcoeffSymb_{\if\relax\detokenize{#1}\relax\else#1\fi}}
\newcommand{\espec}{\sigma}

\newcommand{\specEnthalpy}{h}
\newcommand{\specIntEnergy}{u}
\newcommand{\specVolume}{v}

\newcommand{\residual}[1][]{r_{\if\relax\detokenize{#1}\relax\else\mathrm{#1}\fi}}
\newcommand{\redVolumeFactor}{\mathcal{V}}
\newcommand{\absperm}{\mathbf{K}}
\newcommand{\normperm}[1][]{\mathcal{K}_{\perp\if\relax\detokenize{#1}\relax\else,#1\fi}}
\newcommand{\viscosity}[1][]{\mu\if\relax\detokenize{#1}\relax\else_{#1}\fi}
\newcommand{\thermconducivity}[1][]{\kappa\if\relax\detokenize{#1}\relax\else_{#1}\fi}
\newcommand{\normconductivity}[1][]{\kappa_{\perp\if\relax\detokenize{#1}\relax\else,#1\fi}}
\newcommand{\relperm}[1][]{k_{r\if\relax\detokenize{#1}\relax\else#1\fi}}
\newcommand{\mobility}[1][]{%
  \frac{\relperm[#1]}{\viscosity[#1]}
}
\newcommand{\massmobility}[1][]{\lambda_{\if\relax\detokenize{#1}\relax\else#1\fi}}
\newcommand{\advflux}[1][]{\boldsymbol{v}_{\if\relax\detokenize{#1}\relax\else#1\fi}}
\newcommand{\heatflux}[1][]{\boldsymbol{q}_{\if\relax\detokenize{#1}\relax\else#1\fi}}
\newcommand{\intfadvflux}[1][]{\nu_{\if\relax\detokenize{#1}\relax\else#1\fi}}
\newcommand{\intfheatflux}[1][]{q_{\if\relax\detokenize{#1}\relax\else#1\fi}}
\newcommand{\darcyflux}[1][]{\mobility[#1] \absperm \grad{p}}

\newcommand{\porepy}{\textit{PorePy} }

\DeclareMathOperator*{\argmin}{\text{arg min}}

\NewDocumentCommand{\cop}{O{} o}{
    \operatorname{%
        \if\relax\detokenize{#1}\relax
            \mathcal{F}
        \else
            \mathcal{#1}%
        \fi
    }%
    \IfValueT{#2}{_{#2}}
}
\NewDocumentCommand{\dop}{O{} o}{
    \operatorname{%
        \if\relax\detokenize{#1}\relax
            \mathrm{F}
        \else
            \mathrm{#1}%
        \fi
    }%
    \IfValueT{#2}{_{#2}}
}
\NewDocumentCommand{\cfcop}{O{}}{
    \cop[F][#1]
}
\NewDocumentCommand{\cfdop}{O{}}{
    \dop[F][#1]
}
\NewDocumentCommand{\lecop}{O{\espec}}{
    \cop[G][#1]
}
\NewDocumentCommand{\ledop}{O{\espec}}{
    \dop[G][#1]
}
\NewDocumentCommand{\flcop}{O{\espec}}{
    \cop[S][#1]
}
\NewDocumentCommand{\fldop}{O{\espec}}{
    \dop[S][#1]
}
\NewDocumentCommand{\jaccop}{O{}}{
    \cop[J][#1]
}
\NewDocumentCommand{\jacdop}{O{}}{
    \dop[J][#1]
}

\tikzstyle{arrow} = [thick,->,>=stealth]
\tikzstyle{process} = [rectangle, minimum width=1.5cm, minimum height=0.5cm, text width=1.5cm, text centered, draw=black]
\tikzstyle{decision} = [diamond, minimum width=0.1cm, minimum height=0.1cm, aspect=2, text centered, draw=black, fill=black!5]
\tikzstyle{optprocess} = [rectangle, minimum width=1.5cm, minimum height=0.5cm, text width=1.5cm, text centered, draw=black, dashed]
\tikzstyle{every node}=[font=\tiny]

\def\WFfill{\par 
    \ifx\parshape\WF@fudgeparshape 
    \nobreak 
    \ifnum\c@WF@wrappedlines>\@ne 
    \advance\c@WF@wrappedlines\m@ne 
    \vskip\c@WF@wrappedlines\baselineskip 
    \global\c@WF@wrappedlines\z@ 
    \fi 
    \allowbreak 
    \WF@finale 
    \fi 
}

\makeatother
\newcommand{\correspondant}[1]{\textit{Corresponding author:} #1}
\providecommand{\keywords}[1]{\textbf{\textit{Keywords---}} #1}

\newcommand{\simcase}[1]{\mbox{\texttt{#1}}}
\newcommand{\myunit}[1]{[\unit{#1}]}

\crefname{algorithm}{algorithm}{algorithms}
\Crefname{algorithm}{Algorithm}{Algorithms}

\title{Isochoric thermodynamic preconditioning for resolving mechanically induced phase change in fractured porous media}
\date{\today}
\author[1]{Veljko Lipovac \orcidlink{0000-0002-9442-0879}}
\author[1]{Eirik Keilegavlen \orcidlink{0000-0002-0333-9507}}
\author[1]{Inga Berre \orcidlink{0000-0002-0212-7959}}

\affil[1]{Center for Modeling of Coupled Subsurface Dynamics \\ Department of Mathematics, University of Bergen, Bergen, Norway}

\begin{document}

\maketitle

\begin{abstract}
    Rapid pore-volume changes can trigger phase transitions on time scales shorter than characteristic transport times and may therefore be skipped by conventional nonlinear solves and adaptive time stepping.
We present a persistent-variable framework for thermal compositional flow in fractured porous media that introduces specific volume as an independent transport variable.
Starting from a fully coupled flow–transport–equilibrium system, we derive volume-based models and recover classical pressure-based formulations by eliminating local thermodynamic variables.
To resolve abrupt fracture opening, we introduce a nonlinear preconditioner that assumes instantaneous free expansion and resolves the resulting state through an isochoric equilibrium calculation before advancing the coupled transport problem.
The preconditioner applies to both volume- and pressure-based formulations.

In the studied fracture-opening cases, unpreconditioned simulations miss transient vaporization, whereas the preconditioned models resolve it.
Across the investigated aperture range, larger openings produce monotonic increases in peak gas content, expansion-induced cooling, and durations of gas and pressure transients.
Thermal effects alter the phase evolution but have otherwise minor influence on the overall transient duration.
Within the proof-of-concept setting, fracture opening generates substantial transient pressure reductions, indicating that geomechanical feedback may become important in fully coupled applications.
Pressure–enthalpy and volume–temperature formulations recover indistinguishable physical solutions but exhibit different nonlinear robustness, with the pressure–enthalpy formulation proving more robust in some recompression-dominated cases.
These results show that the equilibrium specification controls the numerical properties of the nonlinear problem rather than the recovered physical response, while isochoric preconditioning connects the nonlinear initialization directly to the underlying thermodynamics.
\end{abstract}


\correspondant{}{Veljko Lipovac \texttt{veljko.lipovac@uib.no}}

\section{Introduction}\label{sec:introduction}
Pressure has historically been the dominant mechanical state variable in reservoir simulation.
From the earliest black-oil models to modern compositional formulations \cite{lake1989-book,trangenstein1989-1,trangenstein1989-2}, governing equations have almost exclusively been formulated in terms of pressure along with quantities describing the transported fluid composition \cite{voskov2012}.
This choice is natural from both a physical and a mathematical perspective.
Pressure acts as the driving force for flow through Darcy's law, while pressure equations exhibit favorable elliptic properties \cite{leveque2002-book} and have formed the basis for decades of developments in numerical discretization, linear solvers, and preconditioning techniques \cite{aavatsmark2002,klausen2008,friis2009,roy2020,wallis1985,zabegaev2026}.
Consequently, pressure-based formulations remain the prevailing approach in contemporary reservoir simulators.

The choice of thermodynamic state variables has followed a similar evolution.
Early equilibrium calculations were formulated in terms of pressure and temperature \cite{michelsen1982-1,michelsen1982-2,rachford1952}, reflecting the natural intensive variables of classical thermodynamics.
As thermal processes gained increasing importance in enhanced oil recovery, geothermal energy production, and other subsurface applications, however, limitations of the pressure--temperature description became apparent \cite{zhu2015}.
Within the two-phase region, temperature remains constant during evaporation or condensation at fixed pressure, whereas thermodynamic quantities such as enthalpy evolve continuously with the progressing phase transition \cite{zhu2014,zhu2016,michelsen1987}.
As a result, pressure--enthalpy formulations emerged as the preferred description for thermal compositional flow, providing a unique characterization of equilibrium states throughout the two-phase region \cite{weiss2014,wang2020,garipov2018,faust1979-1,hayba1994,hu2020}.
Today, enthalpy-based formulations have become the standard approach for thermal reservoir simulation and are employed by several modern simulation frameworks \cite{openpm,voskov2024,mrst-1,settgast2024}.

The history of volume-based formulations has developed quite differently.
Rather than replacing pressure only within the thermodynamic calculations, \citet{acs1985} proposed eliminating the pressure equation altogether by introducing fluid specific volume through the underlying equation of state.
The governing equations then consist solely of transport equations for conserved quantities together with an algebraic volume-balance relation enforcing consistency between the fluid volume predicted by the equation of state and the pore volume provided by the porous medium.
From a thermodynamic perspective, this formulation is particularly attractive because pressure becomes a dependent quantity recovered directly from the equation of state.
More recent studies have employed isochoric or density-based equilibrium calculations in more specialized settings, including thermal cases \cite{polivka2014,li2022,ritschel2019,ibrahim2010}.
Although these approaches remain relatively uncommon, they indicate that volume-based thermodynamic formulations are feasible and provide a natural framework for problems in which pore-volume changes are prescribed.

A second and largely independent motivation for volume-based formulations arises from geomechanics.
Coupled flow and geomechanical models have demonstrated that deformation of the porous medium significantly affects porosity, permeability, fracture apertures, and consequently fluid flow \cite{scott2020,grunwald2022,cusini2021,coussy2003-book,berre2020}.
These couplings are important for engineering applications such as geothermal energy production, carbon sequestration, hydraulic stimulation, and induced seismicity \cite{ucar2017,keilegavlen2021,stefansson2021}, as well as natural earthquake dynamics.
From the perspective of the fluid, however, geomechanics introduces a fundamentally different type of coupling than pressure-driven transport.
Mechanical disturbances propagate through the porous medium on time scales that are typically much shorter than those governing fluid flow and transport.
Consequently, changes in the available pore volume may appear nearly instantaneous when viewed on the characteristic time scale of the fluid.

Rapid dilation of fluid-filled faults and fractures has long been associated with substantial transient reductions in fluid pressure.
First-order piston models predict that co-seismic opening of dilational jogs can drive fluid vaporization and generate low-pressure sinks within fault networks \cite{weatherley2013}.
Direct measurements during dynamic rupture subsequently demonstrated that fault dilation can reduce on-fault fluid pressure to vapor pressure, followed by gradual fluid recharge from the surrounding rock \cite{brantut2020}.
More recently, coupled calculations of fault opening and post-seismic heat and mass transport have shown that the resulting vapor front may propagate beyond the dilational fracture and persist substantially longer than the initial pressure disturbance \cite{alfaro2024}.

These studies establish the physical relevance of dilation-induced phase change but treat the rapid pore-volume alteration either through a reduced thermomechanical model, an undrained pressure estimate, or a separately computed post-opening state supplied to a subsequent transport simulation.
A general formulation in which abrupt pore-volume changes are transmitted directly to local compositional equilibrium while remaining compatible with both pressure- and volume-based transport models has until now not been developed.

The present work builds upon this observation by extending volume-based compositional formulations to a persistent-variable description employing isochoric equilibrium calculations \cite{lipovac2023,lipovac2025}.
Rather than treating fluid volume solely as an algebraic closure variable, we introduce specific volume as an independent state variable whose evolution is governed simultaneously by fluid transport and externally imposed changes in pore volume.
The resulting formulation naturally gives rise to volume--temperature and volume--internal-energy equilibrium calculations while retaining compatibility with conventional pressure--enthalpy descriptions.

As a first step towards fully coupled flow and geomechanics simulations, we consider a simplified setting in which geomechanical effects are represented through prescribed discontinuous changes in fracture aperture.
From the perspective of the fluid, these changes correspond to abrupt variations in the available pore volume and, therefore, provide a suitable model for investigating mechanically induced phase transitions.

We establish isochoric equilibrium calculations as a physically consistent framework for resolving mechanically induced phase transitions resulting from rapid pore-volume changes, providing new insights into the interaction between fracture deformation and fluid phase behavior.
Such a process is well approximated as a free expansion, for which specific volume and internal energy constitute the natural thermodynamic state variables.
In contrast to approaches that prescribe a separately calculated post-opening thermodynamic state as the initial condition for a subsequent flow simulation, the present formulation incorporates the rapid thermodynamic transition as a nonlinear preconditioning step within the time-dependent flow problem.
The same equilibrium transition can therefore initialize pressure- and volume-based transport formulations, allowing the thermodynamic representation and the global choice of primary variables to be selected independently.

The numerical examples reveal that mechanically induced free expansion may generate pronounced transient pressure reductions together with temporary vaporization, even in the absence of external heat transfer.
These results indicate that abrupt changes in fracture aperture can substantially alter local thermodynamic conditions and suggest that the resulting pressure excursions may significantly influence the subsequent mechanical response of fractured porous media.
Although the formulation is derived for general multicomponent mixtures, the numerical examples focus on pure water in order to isolate the effects of mechanically induced phase change.
Consequently, the present work establishes a foundation for future fully coupled flow--geomechanics formulations in which fluid phase behavior and mechanical deformation evolve simultaneously.

\section{Mathematical model}\label{sec:model}
We derive the coupled system of flow, transport, and local equilibrium equations for non-reactive multiphase multicomponent fluid mixtures, neglecting gravity, capillary effects, and molecular diffusion.
Under the assumption of local thermodynamic equilibrium, all fluid phases share a common pressure and temperature, while phase separation is assumed to occur instantaneously.
The governing equations for flow and transport are presented in \Cref{subsec:governing}, followed by the formulation of the local equilibrium problem in \Cref{subsec:local-equilibrium}.
The mixed-dimensional discrete fracture model employed in our \porepy \cite{keilegavlen2020} implementation is described in \Cref{subsec:md-model}, while \Cref{subsec:constitutive} introduces the equation-of-state-based fluid model together with the constitutive relations required by the mixed-dimensional formulation.
Finally, \Cref{subsec:coupling} establishes the coupling between flow and thermodynamic equilibrium through independent volume and energy variables.
The resulting formulation naturally gives rise to volume-based flow models and demonstrates that classical pressure-based formulations are recovered as reduced forms of the proposed framework.

\subsection{Conservation equations}\label{subsec:governing}
Consider a fluid of $\numof{C}$ components, e.g., water, and a fixed number of possible phases $\numof{P}$, e.g., liquid and gas.
We employ a compositional formulation for flow and transport based on the overall mass fractions $\ofrac$ associated with component $\cidx\in\{1,\dots,\numof{C}\}$.
The fraction of mass contained in phase $\pidx\in\{1,\dots,\numof{P}\}$ is denoted by $\pfrac$, and the fraction of pore volume occupied by the phase is given by the saturation $\sat$.
The partial mass fraction of component $\cidx$ in phase $\pidx$ is given by $\cpfrac$.
All fractions fulfill the unity constraint, i.e., 
\begin{equation}\label{equ:unity-fractions}
    \sum\limits_{\alpha}\xi_{\alpha} = 1,
\end{equation}
for $\xi_{\ast}\in\{\ofrac[\ast], \pfrac[\ast], \sat[\ast], \cpfrac[\ast 1], \dots, \cpfrac[\ast\numof{P}]\}$.
Consequently, without loss of generality, the first fraction in each family can be eliminated as an independent degree of freedom using \Cref{equ:unity-fractions}.

Under the assumptions of negligible gravity and equal pressure across all phases, the flux of each phase is described by Darcy's law in the form
\begin{equation}\label{equ:darcy-flux}
    \advflux[\pidx] = -\darcyflux[\pidx],
\end{equation}
where $\relperm[\pidx]$ denotes the relative permeability of phase $\pidx$, $\viscosity[\pidx]$ its viscosity, and $\absperm$ the absolute permeability tensor.
The pressure $p$ acts as the sole driving force for mass transport.

Using \Cref{equ:darcy-flux}, the pressure equation is formulated as the balance of total fluid mass,
\begin{equation}\label{equ:pressure-equation}
    \pdv{}{t}\phi\rho
    - \div{\left(
        \massmobility \absperm \grad{p}
    \right)}
    =
    b ,
\end{equation}
where $\phi$ denotes the porosity and  $\rho$ the total fluid density.
The total mass mobility $\massmobility$ is defined by
\begin{equation}\label{def:total-mass-mobility}
    \massmobility = \sum_{\pridx}\rho_{\pridx}\mobility[\pridx],
\end{equation}
where $\rho_{\pridx}$ denotes the density of phase $\pridx$.
The transport of each independent fluid component, $\cidx\in\{2,\dots,\numof{C}\}$, is described by
\begin{equation}\label{equ:component-balance}
    \pdv{}{t}\phi\rho\ofrac
    - \div{\left(
        \massmobility[\cidx] \absperm \grad{p}
    \right)}
    =
    b_{\cidx}.
\end{equation}
The corresponding component mass mobility is given by
\begin{equation}\label{def:component-mass-mobility}
    \massmobility[\cidx] = \sum_{\pridx}\rho_{\pridx}\cpfrac[\cidx\pridx]\mobility[\pridx].
\end{equation}
For consistency, the source terms must satisfy
\begin{equation}
    b = \sum_{\cridx} b_{\cridx}.
\end{equation}
Summing \Cref{equ:component-balance} over all components $\cidx$ recovers the pressure equation \eqref{equ:pressure-equation} by virtue of the unity constraint \eqref{equ:unity-fractions}.

For thermal simulations, we additionally consider the balance of total energy
\begin{equation}\label{equ:energy-balance}
    \pdv{}{t}\left(
    \phi\rho\specIntEnergy + (1 - \phi)\rho_s \specIntEnergy_s
    \right)- \div{\left(
        \massmobility[\specEnthalpy] \absperm \grad{p}
        +
        \thermconducivity[\text{eff}] \grad{T}
    \right)}
    =
    b_m + b_h,
\end{equation}
where $\specIntEnergy$ denotes the specific internal energy of the fluid, $\specIntEnergy_s$ that of the porous rock, $T$ the common temperature of all phases, $b_m$ the energy source associated with the mass source $b$, and $b_h$ additional heat sources.
The diffusive heat flux is modeled by Fourier's law, where $\thermconducivity[\text{eff}]$ denotes the total effective thermal conductivity.
The enthalpy mobility $\massmobility[\specIntEnergy]$ is defined as
\begin{equation}\label{def:enthalpy-mobility}
    \massmobility[\specEnthalpy] = \sum_{\pridx}\rho_{\pridx}\specEnthalpy_{\pridx}\mobility[\pridx],
\end{equation}
where $\specEnthalpy_{\pridx}$ denotes the specific enthalpy of phase $\pridx$.

\Cref{equ:pressure-equation,equ:component-balance,equ:energy-balance} constitute a system of $\numof{C} + 1$ partial differential equations describing mass transport, the evolution of the mechanical state of the fluid, and the evolution of its thermal state.
Accordingly, the formulation requires $\numof{C} - 1$ independent overall mass fractions, one variable describing the mechanical state, and one variable describing the thermal state.
When these variables are chosen appropriately, phase separation can be resolved locally by minimizing a suitable thermodynamic potential.
The primary transport variables thereby determine the secondary thermodynamic variables, including the saturations $\sat$, phase fractions $\pfrac$, and partial mass fractions $\cpfrac$, which are treated as locally dependent variables.
This naturally leads to the formulation of the local equilibrium problem.

\subsection{Local equilibrium model}\label{subsec:local-equilibrium}
The local equilibrium problem is a thermodynamic phase-separation problem.
A target fluid state is characterized in terms of its mass, mechanical state, and thermal state, while the solution consists of the distribution of mass among the phases and, where applicable, the values of conjugate mechanical or thermal state functions.

For example, given $p$, $T$, and the overall mass fractions $\ofrac$, equilibrium is attained when the Gibbs energy of the fluid is minimized, with the corresponding minimizing values of $\pfrac$ and $\cpfrac$ constituting the solution.
The other equilibrium specifications considered in this work are those at fixed $p$ and $\specEnthalpy$, where the conjugate temperature $T$ becomes an additional degree of freedom; at fixed specific volume $\specVolume$ and $T$, where the conjugate pressure $p$ is unknown at equilibrium; and at fixed specific internal energy $\specIntEnergy$ and specific volume $\specVolume$, where both conjugate variables ($p$ and $T$) become unknowns.
All equilibrium specifications share the common feature that they minimize an appropriate thermodynamic potential.
For an overview of the state-function-based optimization approach to phase equilibria, we refer to \citet{michelsen1999}.

A particular difficulty of phase-equilibrium problems is that the physical presence of phases is not known a priori.
Consequently, the set of physically meaningful variables $\pfrac$ and $\cpfrac$ is initially unknown.
This issue can be resolved using either phase stability checks \cite{michelsen1982-1,michelsen1982-2} or a persistent-variable formulation that employs the full set of unknowns and introduces additional non-negativity constraints into the optimization problem \cite{gharbia2021}.
We employ the persistent-variable formulation for all equilibrium specifications.
It introduces a local, semi-smooth system of algebraic equations representing first-order necessary conditions for optimality.
In the following, we summarize only the systems required for the present work and refer to previous publications for a detailed derivation \cite{lipovac2023,lipovac2025}.

Let $\ecpfrac$ denote the extended partial fractions satisfying
\begin{subequations}\label{def:ext-frac}
    \begin{align}
        \sum\limits_{\cridx}\ecpfrac[\cridx\pidx] &
        \begin{cases}
            = 1 & \text{ if } \pfrac > 0,\\
            < 1 & \text{ if } \pfrac = 0,
        \end{cases} \label{def:ext-frac-unity}\\
        \cpfrac &= \frac{\ecpfrac}{\sum_{\cridx=1}^{\numof{C}} \ecpfrac[\cridx\pidx]}.
        \label{def:ext-frac-par-frac}
    \end{align}
\end{subequations}
For any  $\espec\in\{(p,T), (p,\specEnthalpy), (\specVolume, T), (\specIntEnergy, \specVolume)\}$, we can now write the thermodynamic variables as
\begin{equation}\label{def:secvars}
    \begin{aligned}
        \secvars_{pT} &= \left[\pfrac[2],\dots,\pfrac[\numof{P}],\ecpfrac[11],\dots,\ecpfrac[\numof{C}\numof{P}]\right], \\
        \secvars_{p\specEnthalpy} &= \left[T\right] \times \secvars_{pT}, \\
        \secvars_{\specVolume T} &= \left[p\right] \times \secvars_{pT}, \\
        \secvars_{\specIntEnergy\specVolume} &= \left[p, T\right] \times \secvars_{pT}. \\
    \end{aligned}
\end{equation}
The optimization problem is then written as:
\begin{subequations}\label{problem:equilibrium}
    \begin{align}
        \secvars_{\espec}^{\star} &= \argmin_{\secvars_{\espec}} f_{\espec} \left(\secvars_{\espec}\right),
        \label{problem:equilibrium-min} \\[1ex]
        \text{such that:~~}0 &= \residual[\ofrac](\secvars_{\espec}) = \ofrac - \sum\limits_{\pridx=1}^{\numof{P}} \pfrac[\pridx]\cpfrac[\cidx\pridx], ~ \cidx \in \{2, \dots, \numof{C}\},
        \label{problem:equilibrium-equality-constraints} \\[1ex]
        0 &\leq g_{\pidx}(\secvars_{\espec}) = \pfrac, ~\forall \pidx,
        \label{problem:equilibrium-inequality-constraints}
    \end{align}
\end{subequations}
where $f_{\espec}$ denotes the corresponding thermodynamic potential \cite{michelsen1999,lipovac2023}.
\Cref{problem:equilibrium-equality-constraints} constrains the phase-separated mass to the prescribed overall mass fractions $\ofrac$.
\Cref{problem:equilibrium-inequality-constraints} constrains the phase fractions to the interval $[0,1]$, noting that $\pfrac[1]$ is expressed using \Cref{equ:unity-fractions}.
It is also the key difference from the standard formulation, where no inequality constraints are used, and phase stability tests determine the actual shape of $\secvars_{\espec}$.
In contrast, the persistent-variable formulation employs a fixed set of variables, allowing phase fractions to attain the value of zero.

The first-order necessary optimality conditions for Problem \eqref{problem:equilibrium} give rise to a Karush–Kuhn–Tucker (KKT) system.
We write the system in semi-smooth form using the element-wise $\min$-function as
\begin{equation}\label{equ:local-equilibrium-general}
    \lecop(\primvars_{\espec}, \secvars_{\espec}) = \begin{bmatrix}
        \residual[\espec](\secvars_{\espec}) \\
        \residual[\fugcoeffSymb](\secvars_{\espec}) \\
        \residual[\ofracSymb](\secvars_{\espec}) \\ 
        \min{\left(g(\secvars_{\espec}),\lambda(\secvars_{\espec}) \right)}
    \end{bmatrix} = 0.
\end{equation}
Equilibrium specifications $\espec$ augment the system with the block $\residual[\espec]$, which equates the target state function value, replacing $p$ or $T$ with the respective extensive property of the fluid.
Specifically,
\begin{subequations}\label{def:espec-residuals}
    \begin{align}
        \residual[pT] &= \emptyset\\
        \residual[p\specEnthalpy] &= \sum_{\pridx=1}^{\numof{P}} \pfrac[\pridx] \specEnthalpy_{\pridx} - \specEnthalpy,\\
        \residual[\specVolume T] &= \sum_{\pridx=1}^{\numof{P}} \pfrac[\pridx] \specVolume_{\pridx} - \specVolume,\\
        \residual[\specIntEnergy\specVolume] & = \begin{bmatrix}
            \sum_{\pridx=1}^{\numof{P}} \pfrac[\pridx] \specIntEnergy_{\pridx} - \specIntEnergy \\
            \sum_{\pridx=1}^{\numof{P}} \pfrac[\pridx] \specVolume_{\pridx} - \specVolume
        \end{bmatrix}.
    \end{align}
\end{subequations}
The block $\residual[\fugcoeffSymb]$ contains isofugacity equations of the form
\begin{equation}\label{equ:isofugacity-equations}
    \fugcoeff \ecpfrac - \fugcoeff[\cidx 1]\ecpfrac[\cidx 1] = 0,~ \pidx\in\{2,\dots,\numof{P}\}, \forall \cidx.
\end{equation}
The row block $\residual[\ofracSymb]$ contains the residuals associated with the equality constraints \eqref{problem:equilibrium-equality-constraints}.
As a consequence of the inequality constraints \eqref{problem:equilibrium-inequality-constraints}, $\lecop$ contains a semi-smooth row block.
The vector $\lambda$ is defined as
\begin{equation}\label{def:equilibrium-lagrange-multipliers}
    \lambda_{\pidx}(\secvars_{\espec}) = 1 - \sum_{\cridx=1}^{\numof{C}}\ecpfrac[\cridx\pidx].
\end{equation}

The blocks $\residual[\ofracSymb]$, $\residual[\fugcoeffSymb]$, and $\min$ constitute the basic  $pT$-equilibrium system, which forms the core of all remaining specifications.
Isofugacity \eqref{equ:isofugacity-equations} is imposed for all phases, including physically absent ones.
This, together with the expression for $\lambda$ \eqref{def:equilibrium-lagrange-multipliers}, ensures a well-defined problem with a unique solution \cite{gharbia2021}.

Since the equilibrium problem admits a unique solution for every feasible target state, we can postulate a solution map
\begin{equation}\label{def:flash-solution-map}
    \flcop: \primvars_{\espec} \rightarrow \secvars_{\espec},
\end{equation}
where $\secvars_{\espec}$ is as in \Cref{def:secvars} and $\primvars_{\espec}$ contains the target state
\begin{subequations}\label{def:primvars}
    \begin{align}
        \primvars_0 &= [\ofrac[2], \dots, \ofrac[\numof{C}]],\\
        \primvars_{pT} &= [p, T] \times \primvars_0,\\
        \primvars_{p\specEnthalpy} &= [p, \specEnthalpy] \times \primvars_0,\\
        \primvars_{\specVolume T} &= [\specVolume, T] \times \primvars_0,\\
        \primvars_{\specIntEnergy\specVolume} &= [\specIntEnergy, \specVolume] \times \primvars_0.
    \end{align}
\end{subequations}
Finally, phase fractions and saturations are related through
\begin{equation}\label{equ:phase-fraction-relation}
    \rho \pfrac = \rho_{\pidx}\sat.
\end{equation}
Thus, saturations are obtained directly as a byproduct of the equilibrium calculation.
For notational convenience, we include the independent saturations in $\secvars_{\espec}$ and the corresponding relations \eqref{equ:phase-fraction-relation} among the blocks of $\lecop$, without introducing additional symbols.
This completes the thermodynamic state description in a form directly compatible with the flow and transport model.

\subsection{Mixed-dimensional model}\label{subsec:md-model}
Porous media in the subsurface are often characterized by heterogeneous features with large size-to-width ratios, most notably fractures.
Various approaches exist for modeling and discretizing such media; see, e.g., \citet{berre2019}.
We employ a mixed-dimensional formulation, where fractures are represented as lower-dimensional subdomains immersed in the surrounding rock \cite{alboin2000,alboin2002,martin2005,boon2021}.
These lower-dimensional subdomains are obtained by dimensional reduction through integration over the fracture width, resulting in equations and variables defined on each subdomain and coupled through internal boundary conditions and source terms.
The coupling is defined for each pair consisting of a higher-dimensional and a lower-dimensional neighboring subdomain, where the lower-dimensional subdomain coincides geometrically with an internal boundary of the higher-dimensional one.
Inter-dimensional fluxes of mass and energy are defined on interfaces and enter the model through internal boundary conditions and source terms, respectively.
We have extended the original formulations of \citet{keilegavlen2020} and \citet{stefansson2024} to the general thermal compositional setting in previous work \cite{duran2025,oguntola2025,oguntola2026}; here, we provide only a summary.

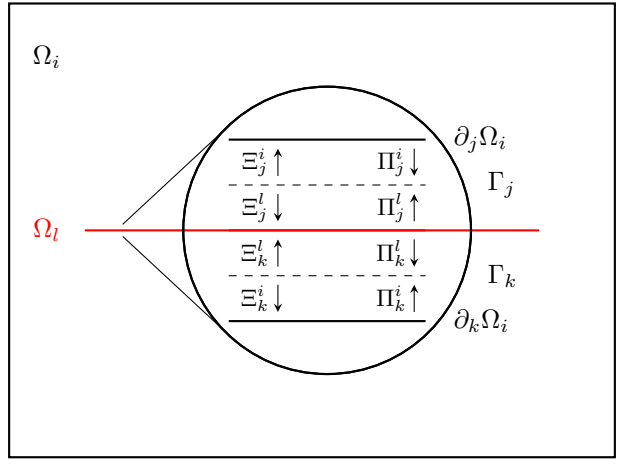
\begin{figure}
\centering
\begin{tikzpicture}[>=stealth]
\def\d{0.6} 
\def\R{1.9} 
\pgfmathsetmacro{\eps}{0.2*\d}

\node[
    rectangle,
    thick,
    draw,
    minimum width=8cm,
    minimum height=6cm
] (domain) at (2,0) {};

\node at (-1.5,2.3) {\normalsize $\Omega_i$};

\draw[red,thick] (-1,0) -- (5,0);
\node[red] at (-1.5,0) {\normalsize $\Omega_l$};

\coordinate (C) at (2.2,0);
\draw[thick] (C) circle (\R);

\draw[thin] (-0.5,0.08) -- ($(C)+(-1.2,1.475)$);
\draw[thin] (-0.5,-0.08) -- ($(C)+(-1.2,-1.475)$);

\begin{scope}

\clip (C) circle (\R);

\def\xXi{-0.65}
\def\xPi{1.15}

\draw[thick] ($(C)+(-1.3, 2*\d)$) -- ($(C)+( 1.3, 2*\d)$);
\draw[dashed] ($(C)+(-1.3, \d)$) -- ($(C)+( 1.3, \d)$);
\draw[red,thick] ($(C)+(-1.3,0)$) -- ($(C)+( 1.3,0)$);
\draw[dashed] ($(C)+(-1.3,-\d)$) -- ($(C)+( 1.3,-\d)$);
\draw[thick] ($(C)+(-1.3,-2*\d)$) -- ($(C)+( 1.3,-2*\d)$);


\node at ($(C)+(\xXi-0.3,1.5*\d)$) {\small $\Xi^i_j$};
\draw[->] ($(C)+(\xXi,\d+\eps)$) -- ($(C)+(\xXi,2*\d-\eps)$);

\node at ($(C)+(\xXi-0.3,0.5*\d)$) {\small $\Xi^l_j$};
\draw[->] ($(C)+(\xXi,\d-\eps)$) -- ($(C)+(\xXi,\eps)$);

\node at ($(C)+(\xXi-0.3,-0.5*\d)$) {\small $\Xi^l_k$};
\draw[->] ($(C)+(\xXi,-\d+\eps)$) -- ($(C)+(\xXi,-\eps)$);

\node at ($(C)+(\xXi-0.3,-1.5*\d)$) {\small $\Xi^i_k$};
\draw[->] ($(C)+(\xXi,-\d-\eps)$) -- ($(C)+(\xXi,-2*\d+\eps)$);


\node at ($(C)+(\xPi-0.3,1.5*\d)$) {\small $\Pi^i_j$};
\draw[->] ($(C)+(\xPi,2*\d-\eps)$) -- ($(C)+(\xPi,\d+\eps)$);

\node at ($(C)+(\xPi-0.3,0.5*\d)$) {\small $\Pi^l_j$};
\draw[->] ($(C)+(\xPi,\eps)$) -- ($(C)+(\xPi,\d-\eps)$);

\node at ($(C)+(\xPi-0.3,-0.5*\d)$) {\small $\Pi^l_k$};
\draw[->] ($(C)+(\xPi,-\eps)$) -- ($(C)+(\xPi,-\d+\eps)$);

\node at ($(C)+(\xPi-0.3,-1.5*\d)$) {\small $\Pi^i_k$};
\draw[->] ($(C)+(\xPi,-2*\d+\eps)$) -- ($(C)+(\xPi,-\d-\eps)$);

\end{scope}

\draw[thick] (C) circle (\R);

\node[right] at ($(C)+(1.55,2*\d)$) {\normalsize $\partial_{j}\Omega_i$};
\node[right] at ($(C)+(2.0,\d)$) {\normalsize $\Gamma_j$};
\node[right] at ($(C)+(2.0,-\d)$) {\normalsize $\Gamma_k$};
\node[right] at ($(C)+(1.55,-2*\d)$) {\normalsize $\partial_{k}\Omega_i$};

\end{tikzpicture}
    \caption{
    A two-dimensional porous medium with a single one-dimensional fracture.
    The exploded view shows the internal boundaries (solid, black), the interfaces (dashed, black), and the lower-dimensional neighbor (solid, red).
    }
    \label{fig:md-domain}
\end{figure}

We first introduce the required notation.
For a visual representation of the terminology, see \Cref{fig:md-domain}.
The fractured medium is composed of individual subdomains $\Omega_i\subset\mathbb{R}^{d_i}, d_i\in\{0,1,2,3\}$, where the ambient rock $\Omega_0$ is the only subdomain of highest dimension $d_0$.
Any quantity introduced in \Cref{subsec:governing,subsec:local-equilibrium} has a corresponding representation on subdomain $i$, denoted by the subscript $i$.
For example, $p_i$ denotes the pressure in $\Omega_i$.
For a pair of neighboring subdomains $\Omega_i$ and $\Omega_l$, representing the higher- and lower-dimensional subdomains, respectively, we introduce interfaces $\Gamma_j$ and $\Gamma_k$ on the two sides of $\Omega_l$.
These correspond to the two sides of the internal boundary of $\Omega_i$, denoted by $\partial_j \Omega_i$ and $\partial_k \Omega_i$.
For each subdomain $i$, we define $\hat{S}_i$ as the set of interfaces to higher-dimensional neighbors.
In particular, $\hat{S}_0 = \emptyset$ for the ambient rock, $\lvert \hat{S}_i \rvert = 2$ for immersed fractures of dimension $d_0 - 1$, and $\lvert \hat{S}_i \rvert = 4$ for the dimension $d_0 - 2$ intersection of two crossing fractures.
Points, constituting subdomains of dimension zero, are coupled with a single interface linking the point to the closest cell of the higher-dimensional neighbor.
To handle inter-dimensional fluxes on interfaces, we introduce projections.
The projection from $\Omega_i$ to $\Gamma_j$ is denoted by $\Pi^i_j$, and the reverse map by $\Xi^i_j$.
Finally, we introduce the notion of fracture aperture $a$ and the volume factor $\redVolumeFactor_i = a^{d_0 - d_i}$, which accounts for the reduced volume after dimensional reduction.
Note that $\redVolumeFactor_0 = 1$ holds for the rock matrix.

\Cref{subsec:governing} introduced two types of fluxes: the Darcy flux and the Fourier flux.
These are proportional to the gradients of pressure and temperature, respectively.
We proceed analogously for inter-dimensional fluxes by introducing the Darcy-like advective interface flux on an interface $\Gamma_j$ as
\begin{equation}\label{def:interface-adv-flux}
    \intfadvflux[j] = -\normperm[j]\frac{2}{\Pi^l_j a_l}\left(\Pi^l_j p_l - \Pi^i_j p_i\right),
\end{equation}
where $\normperm[j]$ denotes the normal permeability at the interface between the subdomains.
In contrast to the single-phase formulation presented by \citet{keilegavlen2020}, we omit the viscosity and interpret $\intfadvflux[j]$ as a mobility-free, force-like quantity.
For the inter-dimensional heat flux, we introduce the Fourier-like interface heat flux as
\begin{equation}\label{def:interface-heat-flux}
    \intfheatflux[j] = -\normconductivity[j]\frac{2}{\Pi^l_j a_l}\left(\Pi^l_j T_l - \Pi^i_j T_i\right),
\end{equation}
where $\normconductivity[j]$ denotes the normal thermal conductivity.
Both inter-dimensional fluxes follow the same principle as their fixed-dimensional counterparts: they are driven by pressure and temperature differences between neighboring subdomains and scaled by factors representing rock and fluid properties.

In total, the mixed-dimensional thermal compositional flow model can be written as
\begin{subequations}\label{equ:md-pde-system}
    \begin{align}
        \pdv{}{t}\redVolumeFactor_i\phi_i\rho_i
            + \div{\left(
                \redVolumeFactor_i \massmobility[i] \absperm_i \grad{p_i}
            \right)}
            + \sum\limits_{j\in\hat{S}_i} \Xi^i_j\left(\redVolumeFactor_j \massmobility[j]\intfadvflux[j]\right)
            &=
            r_i, &\text{ in } \Omega_i,
        \label{equ:md-pde-system-pressure}\\
        \pdv{}{t}\redVolumeFactor_i\phi_i\rho_i\ofrac[\cidx i]
        + \div{\left(
            \redVolumeFactor_i
            \massmobility[\cidx i]\absperm_i \grad{p_i}
        \right)}
        + \sum\limits_{j\in\hat{S}_i} \Xi^i_j \left(\redVolumeFactor_j \massmobility[\cidx j]\intfadvflux[j]\right)
        &=
        r_{\cidx i}, &\text{ in } \Omega_i,
        \label{equ:md-pde-system-component}\\
        \pdv{}{t}\left(
        \redVolumeFactor_i \left(\phi_i\rho_i\specIntEnergy_i + (1 - \phi_i)\rho_{si} \specIntEnergy_{si}\right)
        \right)
        + \div{\left(
            \redVolumeFactor_i
            \left(
            \massmobility[\specEnthalpy i] \absperm_i \grad{p_i} + \thermconducivity[\text{eff},i] \grad{T_i}
            \right)
        \right)}
        & &\nonumber \\
        +\sum\limits_{j\in\hat{S}_i} \Xi^i_j
        \left(
        \redVolumeFactor_j\left(\massmobility[\specEnthalpy j]\intfadvflux[j] + \intfheatflux[j]\right)
        \right)&=
        r_{mi} + r_{hi}, &\text{ in } \Omega_i,
        \label{equ:md-pde-system-energy}\\
        \intfadvflux[j] + \normperm[j]\frac{2}{\Pi^l_j a_l}\left(\Pi^l_j p_l - \Pi^i_j p_i\right)
        &= 0, &\text{ on } \Gamma_j,
        \label{equ:md-pde-system-intf-advective}\\
        \intfheatflux[j] + \normconductivity[j]\frac{2}{\Pi^l_j a_l}\left(\Pi^l_j T_l - \Pi^i_j T_i\right)
        &= 0, &\text{ on } \Gamma_j,
        \label{equ:md-pde-system-intf-heat}\\
        \Xi^i_j \left(\redVolumeFactor_j \massmobility[j]\intfadvflux[j]\right)
        - \redVolumeFactor_i\left(\massmobility[i] \absperm_i \grad{p_i}\right)\cdot\mathbf{n}_{ij}
        &= 0, &\text{ on } \partial_j \Omega_i,
        \label{equ:md-pde-system-pressure-bc}\\
        \Xi^i_j \left(\redVolumeFactor_j \massmobility[\cidx j]\intfadvflux[j]\right)
        - \redVolumeFactor_i\left(\massmobility[\cidx i] \absperm_i \grad{p_i}\right)\cdot\mathbf{n}_{ij}
        &= 0, &\text{ on } \partial_j \Omega_i,
        \label{equ:md-pde-system-component-bc}\\
        \Xi^i_j \left(\redVolumeFactor_j \left(\massmobility[\specEnthalpy j]\intfadvflux[j] + \intfheatflux[j]\right)\right)
        - \redVolumeFactor_i\left(\massmobility[\specEnthalpy i] \absperm_i \grad{p_i} + \thermconducivity[\text{eff},i] \grad{T_i}\right)\cdot\mathbf{n}_{ij}
        &= 0, &\text{ on } \partial_j \Omega_i.
        \label{equ:md-pde-system-energy-bc}
    \end{align}
\end{subequations}
\Crefrange{equ:md-pde-system-pressure}{equ:md-pde-system-energy} represent the mixed-dimensional counterparts of the governing equations in \Cref{equ:pressure-equation,equ:component-balance,equ:energy-balance}.
The main differences are the additional source terms resulting from inter-dimensional fluxes and the presence of the volume factor $\redVolumeFactor_i$.
For the matrix $\Omega_0$, these equations reduce to those presented in \Cref{subsec:governing}, since $\redVolumeFactor_0 = 1$ and $\hat{S}_0 = \emptyset$.
The internal boundary conditions \Crefrange{equ:md-pde-system-pressure-bc}{equ:md-pde-system-energy-bc} are imposed for the pressure, component-mass, and energy balance equations, respectively.
The unit normal vector $\mathbf{n}_{ij}$ points outward on the internal boundary $\partial_j \Omega_i$.
Note that $\mathbf{n}_{ij}$ and $\mathbf{n}_{ik}$ have the same orientation but opposite direction, corresponding to the two sides of the lower-dimensional neighbor and their respective interfaces.
\Cref{equ:md-pde-system-intf-advective,equ:md-pde-system-intf-heat} define the interface unknowns $\intfadvflux[j]$ and $\intfheatflux[j]$.
In principle, these equations and unknowns need not be introduced, since they are algebraic expansions: the expressions in \Cref{def:interface-adv-flux,def:interface-heat-flux} could be inserted directly into \Crefrange{equ:md-pde-system-pressure}{equ:md-pde-system-energy}.
However, this is the structure used by the underlying software \porepy \cite{keilegavlen2020}, and we state the full system for transparency.

Finally, the local equilibrium equations in \Cref{equ:local-equilibrium-general} are introduced separately on each subdomain $\Omega_i$.
Since these equations are local algebraic relations between the primary transport variables and the secondary thermodynamic state variables, no structural modification is required in the mixed-dimensional setting.
Fluid properties appearing in $\massmobility[j], \massmobility[\cidx j]$ and $\normconductivity[j]$, however, require EoS-based evaluations on interfaces.
We approximate the corresponding interface quantities using an upwind-type method, choosing values from the upstream side of the respective interface flux.
The discretization is described in \Cref{sec:numerics}.

\subsection{Constitutive laws}\label{subsec:constitutive}
So far, we have introduced two constitutive assumptions: Darcy's law for advective flow and Fourier's law for diffusive heat transport.
Both treat the pressure $p$ and temperature $T$ as potentials driving the corresponding transfer processes, weighted by scalar quantities composed of fluid and rock properties.
For simplicity, we use linear relative permeabilities,
\begin{equation}
    \relperm[\pidx] = \sat[\pidx].
\end{equation}
More complex models, such as Brooks--Corey-type functions \cite{brooks1964}, are possible but introduce additional nonlinearities that are not the focus of this work.
For the effective thermal conductivity, we apply the linear mixing rule
\begin{equation}\label{def:eff-thermal-conductivity}
    \thermconducivity[\text{eff}] = \phi\left(\sum_{\pridx}\sat[\pridx]\thermconducivity[\pridx]\right) + (1 - \phi)\thermconducivity[s],
\end{equation}
where $\thermconducivity$ denotes the thermal conductivity of the individual phases and the rock.
For more elaborate models, we refer to the existing literature; see, e.g., \citet{tong2009}.

In the mixed-dimensional setting, additional quantities are introduced to account for the reduced dimension of the immersed fractures.
One such quantity is the normal permeability $\normperm[j]$, defined as
\begin{equation}\label{def:normal-permeability}
    \normperm[j] = \Pi^l_j\absperm_l.
\end{equation}
As discussed by \citet{stefansson2021}, the normal permeability results from the dimensional reduction of the lower-dimensional domain.
It is therefore chosen on the interface as the projection of the absolute permeability of the corresponding lower-dimensional subdomain.
By analogous reasoning, we define the normal thermal conductivity as
\begin{equation}\label{def:normal-thermal-conductivity}
    \normconductivity[j] = \Pi^l_j\thermconducivity_{eff, l},
\end{equation}
i.e., as the projection of the effective thermal conductivity in the lower-dimensional subdomain onto the interface.
This completes the constitutive description of the flow and transport model.

Rock properties are modeled using simple heuristic laws and reference values, assuming that the rock neither reacts with the fluid nor undergoes any additional thermodynamic state changes.
In particular, we assume constant density $\rho_s$, constant thermal conductivity $\thermconducivity[s]$, and constant specific heat capacity at constant volume $c_{\specVolume, s}$.
The latter assumption allows us to express the specific internal energy of the rock as
\begin{equation}\label{def:energy-rock}
    \specIntEnergy_s = c_{\specVolume, s}\left(T - T_0\right),
\end{equation}
where $T_0$ is a reference temperature provided together with $c_{\specVolume, s}$.

Fluid properties are more involved.
To demonstrate the capabilities of the formulation, we use a sufficiently rich equation-of-state-based fluid model.
For general subsurface fluid mixtures, the cubic Peng--Robinson (PR) EoS is a standard choice \cite{peng1976,lopez2017}.
Since extensions of the PR-EoS are available for evaluating properties of absent phases \cite{gharbia2021,lipovac2023}, it is used in this work.
The proposed formulation itself, however, is not tied to this particular EoS, and more accurate fluid models exist for water-dominated systems \cite{iapws1997,driesner2007}.
For the present general compositional persistent-variable formulation, we restrict ourselves to the extended PR-EoS.

We use a pressure--temperature-based representation of the fluid properties.
Thus, for each phase $\pidx$, any fluid property $\psi_{\pidx}$ is expressed as $\psi_{\pidx} = \psi_{\pidx}(p, T, \cpfrac)$ through functions obtained from the EoS.
These include $\rho_{\pidx}, \specIntEnergy_{\pidx}, \specEnthalpy_{\pidx}$ and $\fugcoeff$.
For an overview of these quantities, we refer to our previous work \cite[Appendix B]{lipovac2023}.
Viscosities and thermal conductivities of the fluid phases are not thermodynamic state functions.
They are transport properties and require additional constitutive assumptions.
In this work, we assume constant values for both liquid and gas phases, which is sufficient for the numerical examples considered in \Cref{sec:examples}.
For simulations with a more persistent gas phase, more advanced models are required to describe the advection of liquid and gas phases accurately \cite{lohrenz1964,cardona2019,guo2001}.

In general, whole-fluid properties appearing in the accumulation terms of \Cref{equ:pressure-equation,equ:energy-balance}, such as density and internal energy, are obtained as weighted sums of phase properties:
\begin{subequations}\label{def:fluid-properties}
    \begin{align}
        \rho &= \sum_{\pridx}\sat[\pridx]\rho_{\pridx}, \label{def:density}\\
        \specIntEnergy &= \sum_{\pridx}\pfrac[\pridx]\specIntEnergy_{\pridx}. \label{def:internal-energy}
    \end{align}
\end{subequations}
We also record two thermodynamic identities used in the following sections.
The first is the definition of fluid specific volume
\begin{equation}\label{def:specific-volume}
    \frac{1}{\rho} = \specVolume = \sum_{\pridx}\pfrac[\pridx]\specVolume_{\pridx}.
\end{equation}
The second is the definition of specific enthalpy
\begin{equation}\label{def:specific-enthalpy}
    \specEnthalpy
    = \sum_{\pridx}\pfrac[\pridx]\specEnthalpy_{\pridx}
    = \sum_{\pridx}\pfrac[\pridx]\specIntEnergy_{\pridx} + p\left( \sum_{\pridx}\pfrac[\pridx]\specVolume_{\pridx}\right)
    =\specIntEnergy + p\specVolume.
\end{equation}
On the one hand, enthalpy is obtained as the phase-fraction-weighted sum of phase enthalpies.
On the other hand, the specific enthalpy of any fluid is the sum of specific internal energy and pressure work.
These identities are used below to formulate thermal compositional models based on different thermodynamic state functions and to employ them explicitly as independent variables in the transport problem.
In this way, the variables $\primvars_{\espec}$ characterize both the thermodynamic fluid state and the phase composition.

\subsection{Volume-based coupling}\label{subsec:coupling}
The preceding subsections introduced the components of the thermal compositional flow model: the mixed-dimensional balance equations, the local equilibrium problem, and the constitutive relations.
We now combine these ingredients into a closed formulation and specify the primary variables used for each equilibrium specification.

The local equilibrium problem separates the thermodynamic variables into two groups.
The first group, denoted by $\primvars_{\espec}$, consists of the available mass and two state functions that determine the thermodynamic state of the fluid.
For every feasible input state $\primvars_{\espec}$, the equilibrium problem provides a unique set of secondary thermodynamic variables $\secvars_{\espec}$, including phase fractions, saturations, and phase compositions.
This relation is given by the solution map $\secvars_{\espec} = \flcop(\primvars_{\espec})$ in \Cref{def:flash-solution-map}.

The governing equations describe the evolution of the primary state $\primvars_{\espec}$, while the local equilibrium equations determine the corresponding secondary variables point-wise.
Using the thermodynamic identities in \Crefrange{def:fluid-properties}{def:specific-enthalpy}, the accumulation terms in the governing equations can be expressed in terms of the selected primary state functions.
Thus, the mixed-dimensional transport problem \eqref{equ:md-pde-system} may be written abstractly as
\begin{equation}\label{def:cf-system}
    \cfcop(\primvars_{\espec}, \secvars_{\espec})
    =
    \cfcop(\primvars_{\espec}, \flcop(\primvars_{\espec}))
    =
    0.
\end{equation}
This notation emphasizes that the transport problem evolves the primary variables, whereas all phase-dependent quantities are obtained locally from equilibrium.

The need to move beyond $pT$-based formulations becomes apparent already for narrow-boiling fluids such as pure water.
At fixed pressure, temperature remains constant during evaporation, while the gas fraction changes continuously from zero to one.
Consequently, temperature alone cannot distinguish intermediate phase configurations in the two-phase region.
Specific enthalpy, by contrast, increases monotonically during the transition and therefore provides a continuous characterization of the thermal state.
This motivates $p\specEnthalpy$-based formulations, where enthalpy is evolved by the energy balance and temperature is recovered locally from equilibrium \cite{weiss2014,wang2020,garipov2018,faust1979-1,hayba1994,hu2020}.
Moreover, fluid enthalpy couples naturally to the energy stored in the rock and therefore forms the energetic interaction variable between fluid and porous medium.

A similar argument motivates volume-based formulations.
The $\specVolume T$-formulation introduced by \citet{acs1985} can be viewed as replacing pressure by fluid specific volume as the mechanical state variable.
Volume-based thermodynamic formulations, in particular those based on Helmholtz energy, are known to provide a complete and robust description of the thermodynamic state space \cite{tillner1998,bell2016,bell2017}.
For porous-media applications, however, an additional physical motivation exists.
The fluid volume must remain compatible with the available pore volume to avoid the creation of nonphysical voids.
This relation becomes particularly important when the porous medium deforms, since changes in pore volume directly alter the thermodynamic state of the contained fluid.
Consequently, when pore volume becomes a dynamic quantity, fluid specific volume provides the natural state variable through which mechanical deformation enters the local equilibrium problem.

\begin{figure}
    \centering
    \includegraphics[width=0.5\textwidth]{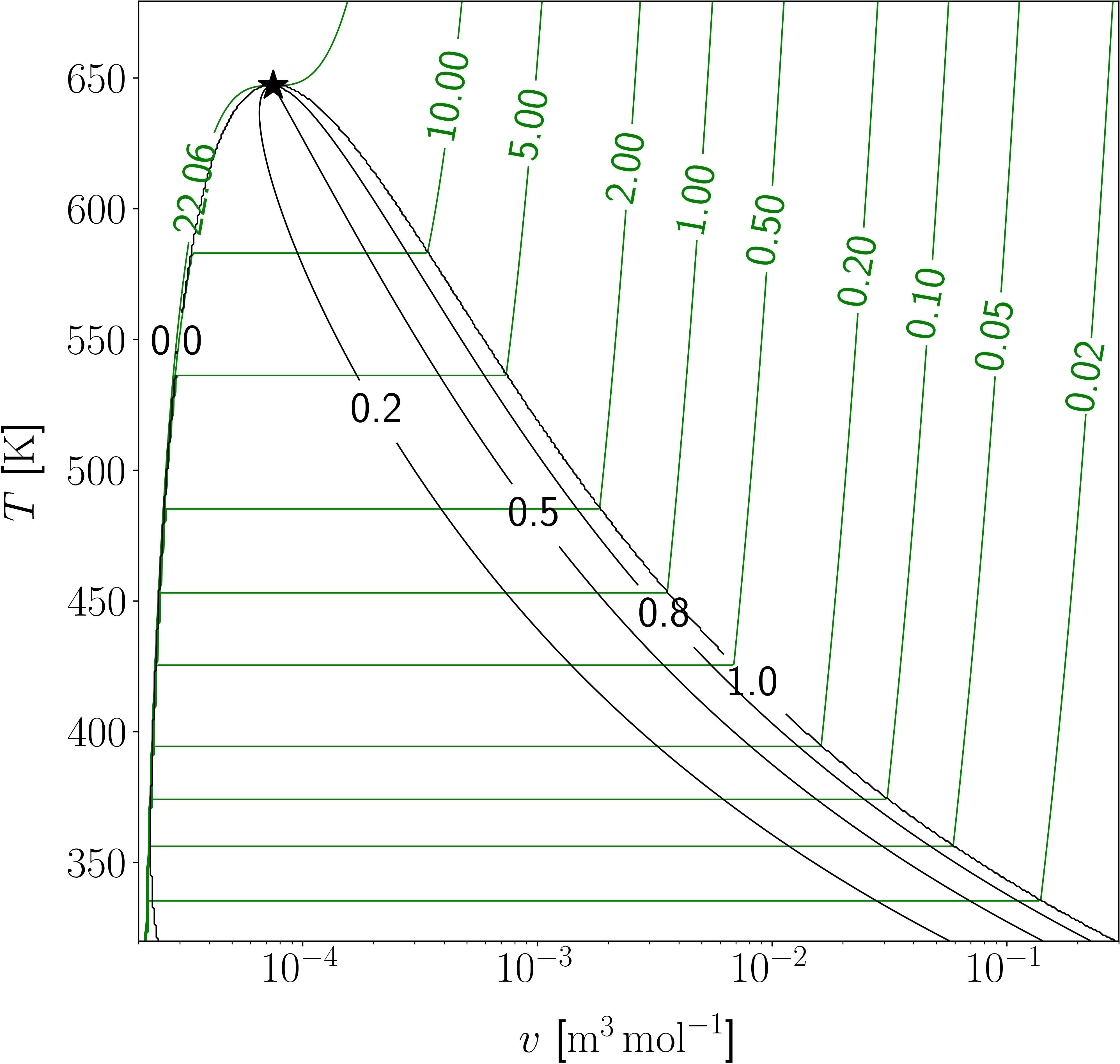}
    \caption{The $\specVolume T$-phase diagram for pure water, computed with the persistent-variable approach using the PR EoS.
    Isolines for gas fraction $\pfrac[]$ are shown in black, and isobaric lines are shown in green.
    Pressure is expressed in \unit{\mega\pascal}.
    The critical point, using the specific volume computed by the EoS, is indicated by the black star.}
    \label{fig:vT-diagram}
\end{figure}

As in the $p\specEnthalpy$-case, isochoric specifications also resolve the two-phase region continuously.
\Cref{fig:vT-diagram} shows the $\specVolume T$-phase diagram for pure water.
Within the two-phase region, the isobars are flat, while the specific volume increases monotonically with increasing gas fraction.
The diagram therefore provides the volume-based analogue of the classical $p\specEnthalpy$-description: the extensive state variable resolves the phase transition, while the conjugate intensive variable is recovered as part of the local equilibrium calculation.

The transition from $pT$-based formulations to specifications involving extensive state variables also clarifies the coupling between transport and equilibrium.
Whenever an extensive state variable, such as $\specEnthalpy$, $\specVolume$, or $\specIntEnergy$, is introduced as a primary transport variable, its conjugate intensive variable is no longer independent.
The local equilibrium system must therefore be augmented by the corresponding residual $\residual[\espec]$, which enforces the definition of the extensive variable as a phase-fraction-weighted sum of phase properties.
Conversely, this residual provides the algebraic closure required by the transport problem.
Thus, the additional equilibrium equations are not merely local thermodynamic constraints; they constitute the closure relations that make extensive-variable flow formulations well posed.

This closes the model as a coupled system for $(\primvars_{\espec},\secvars_{\espec})$.
The transport equations evolve the primary state variables, while the local equilibrium equations determine the associated thermodynamic state and phase configuration.
The next section describes the discretization of this coupled system and the nonlinear solution strategy used for its solution.

\section{Numerical methodology}\label{sec:numerics}
The computational domain is discretized using non-uniform triangular meshes.
In the mixed-dimensional setting, neighboring subdomains are meshed such that the lower-dimensional subdomain and the corresponding interface are geometrically identical to the mesh of the associated internal boundary.

The transport problem, System \eqref{def:cf-system}, is discretized using a finite-volume method.
All degrees of freedom are represented as cell-centered quantities.
Intra-domain fluxes, namely the Darcy and Fourier fluxes, are discretized on cell faces using the multi-point flux approximation (MPFA) \cite{stefansson2018,nordbotten2020}.
The inter-dimensional fluxes on the interfaces, $\intfadvflux[j]$ and $\intfheatflux[j]$, require no additional spatial discretization, since the coupling between neighboring subdomains is algebraic.
Mass and enthalpy mobilities are approximated on cell faces using first-order upwinding \cite{courant1952}.
The mobility terms appearing in the interface equations, \Cref{equ:md-pde-system-intf-advective,equ:md-pde-system-intf-heat}, are approximated analogously using cell-centered values taken from the upstream side of the respective interface flux.
More precisely, if the inter-dimensional flux points from the higher- to the lower-dimensional domain, the values in the cells adjacent to the internal boundary of the higher-dimensional domain are used.
If it points the other way, the values in the corresponding cells of the lower-dimensional subdomain are used.
The local equilibrium equations, System \eqref{equ:local-equilibrium-general}, are formulated cell-center-wise in every subdomain.

In the remainder of this section, we present the global nonlinear solution strategy in \Cref{subsec:num-method}, and an isochoric nonlinear preconditioner in \Cref{subsec:isochoric-preconditiong}, for resolving fluid phase change due to dynamics on unresolved time scales.

\subsection{Fully coupled nonlinear solver}\label{subsec:num-method}
Time integration is performed using the implicit Euler method, resulting in the fully coupled nonlinear system
\begin{subequations}\label{equ:coupled-system}
    \begin{align}
        \cfdop(\primvars_{\espec}, \secvars_{\espec}) &= 0,
        \label{equ:coupled-system-cf} \\[1ex]
        \ledop(\primvars_{\espec}, \secvars_{\espec}) &= 0,
        \label{equ:coupled-system-le}
    \end{align}
\end{subequations}
which must be solved to advance the solution from $t_n$ to $t_{n+1}$.
The symbols $\cfdop$ and $\ledop$ denote the discrete version of $\cfcop$ and $\lecop$, respectively.

Applying a semi-smooth Newton method yields the linear system
\begin{equation}\label{equ:coupled-system-discretized}
    \begin{bmatrix}
        \jacdop[F,\primvars] & \jacdop[F,\secvars] \\
        \jacdop[\ledop,\primvars] & \jacdop[\ledop,\secvars] 
    \end{bmatrix}
    \begin{bmatrix}
        \Delta \primvars_{\espec}^k \\
        \Delta \secvars_{\espec}^k
    \end{bmatrix}
     = -
     \begin{bmatrix}
         \cfdop \\
         \ledop
     \end{bmatrix},
\end{equation}
where $\jacdop[f, x]$ denotes the partial Jacobian of $f$ with respect to the variables $x$, and $k$ represents the nonlinear iteration.
As the convergence criterion, we employ the $L^2(\Omega)$-norm equation-wise, using relative and absolute tolerances.
I.e., the solution is considered converged if
\begin{equation}\label{equ:convergence-criterion}
    \lVert f(\primvars^k_{\espec}, \secvars^k_{\espec}) \rVert_{L^2(\Omega)}
    \leq \epsilon_a + \epsilon_r
    \lVert f(\primvars^{k-1}_{\espec}, \secvars^{k-1}_{\espec}) \rVert_{L^2(\Omega)},
\end{equation}
where $f$ denotes an individual residual equation of either $\cfdop$ or $\ledop$, and $\epsilon_a$ and $\epsilon_r$ are the respective tolerances.

Since the local equilibrium equations form a block-diagonal subsystem, manifesting in a block-diagonal matrix $\jacdop[\ledop,\secvars]$, we eliminate the secondary thermodynamic variables $\secvars_{\espec}$ through a Schur complement reduction.
The resulting system is given by
\begin{subequations}\label{equ:coupled-system-reduced}
    \begin{align}
        \left(\jacdop[F,\primvars] - \jacdop[F,\secvars] \jacdop[\ledop,\secvars]^{-1} \jacdop[\ledop,\primvars]\right) \Delta \primvars_{\espec}^k
        &= -\cfdop + \jacdop[F,\secvars] \jacdop[\ledop,\secvars]^{-1} \ledop, \label{equ:coupled-system-reduced-dx}\\
        \Delta \secvars_{\espec}^k &=
        \jacdop[\ledop,\secvars]^{-1} \left(-\ledop - \jacdop[\ledop,\primvars]\primvars_{\espec}^k\right). \label{equ:coupled-system-reduced-dy}
    \end{align}
\end{subequations}
We employ inexact equilibrium calculations to accelerate convergence.
Every $m$-th nonlinear iteration, the local equilibrium system is solved for fixed $\primvars_{\espec}$, yielding updated values $\secvars_{\espec}$, such that $\lVert \ledop \rVert \approx 0$.
The local system is solved down to an absolute tolerance $\epsilon_{loc}$, which is higher than the global tolerance $\epsilon_a$ applied to the full system $\left[\cfdop, \ledop\right]^{\top}$.
As a local solver, we employ a non-parametric interior point method (NPIPM) \cite{vu2021}.
If the local equilibrium system is solved down to machine precision in every iteration and the Schur complement reduction is performed, a linear system of the form
\begin{equation}\label{equ:conventional-system-discretized}
    \left(\jacdop[F,\primvars] - \jacdop[F,\secvars] \jacdop[\ledop,\secvars]^{-1} \jacdop[\ledop,\primvars]\right) \Delta \primvars_{\espec}^k
    \approx -\cfdop,
\end{equation}
is obtained, which is essentially the linearized version of \Cref{def:cf-system} using $\flcop$.
In the limiting case of exact equilibrium calculations, the matrix $- \jacdop[F,\secvars] \jacdop[\ledop,\secvars]^{-1} \jacdop[\ledop,\primvars]$ represents the exact thermodynamic derivatives in an implicit form.
Consequently, the Schur complement introduces thermodynamically consistent derivatives into the transport Jacobian without requiring these derivatives to be assembled explicitly.
For a discussion of the inexact approach to the fully coupled problem, we refer to our previous work \cite{lipovac2025}.

To globalize the nonlinear iterations, we employ the trust-region Newton method of \citet{park2022}.
Unlike the original formulation, where trust-region iterations are interrupted whenever a phase change necessitates a switch of primary variables, the persistent-variable formulation requires no such intervention.
All variables and equations remain well defined throughout the nonlinear solve.

To further improve robustness, we introduce an additional equilibration step whenever the Newton update is predicted to change the phase configuration.
This is detected by inspecting the preliminary phase fractions
\begin{equation}\label{equ:phase-fraction-update}
    \pfrac^k = \pfrac^{k-1} + \Delta \pfrac^k.
\end{equation}
If any phase fraction leaves the open interval $(0,1)$ while its previous value lies inside it, or conversely enters $(0,1)$ from one of its boundary values $\{0,1\}$, a phase is assumed to disappear or appear, respectively.
In that case, we evaluate an additional equilibrium calculation and replace the raw Newton update of the secondary variables by
\begin{equation}\label{equ:re-equilibrated}
    \begin{aligned}
        \hat{\secvars}_{\espec}
        &=
        \flcop(\primvars^k_{\espec} + \Delta \primvars^k_{\espec}),\\
        \Delta\hat{\secvars}^k_{\espec},
        &=
        \hat{\secvars}_{\espec} - \secvars^{k-1}_{\espec}.
    \end{aligned}
\end{equation}
The trust-region globalization is then performed using the re-equilibrated update.

This modification is motivated by the observation that phase transitions frequently lead to rejected trust-region iterations.
During the appearance or disappearance of phases, the linearization underlying Newton's method provides only a poor local approximation of the equilibrium manifold, and the corresponding trial step often increases the global residual.
The additional equilibrium calculation projects the trial iterate back onto the manifold of thermodynamically admissible states before globalization is applied, resulting in substantially more reliable trial directions.
Moreover, the additional equilibrium calculation is initialized with the predicted state $\secvars^{k-1}_{\espec} +\Delta\secvars^k_{\espec}$,
making it considerably less expensive than rejecting the global Newton iteration, recomputing the Jacobian, and resolving the linearized system.

Prior to the Schur complement reduction, the variables are linearly scaled by linear right preconditioning.
This scaling is particularly important in the mixed-dimensional setting, where variables associated with different subdomains and interfaces may differ substantially in magnitude.
Throughout the numerical examples, dimensional variables are scaled using constant reference values, while fractional quantities, including $\ofrac,\pfrac,\ecpfrac$ and $\sat$, are  naturally of order $\mathcal{O}(1)$ and require no scaling.
We summarize the complete nonlinear solution strategy in \Cref{alg:global}.

Finally, we emphasize that the choice of eliminated variables in the Schur complement reduction must remain consistent with the thermodynamic definition of primary and secondary variables.
In particular, if extensive state variables are introduced in place of $p$ or $T$, it is generally not advisable to eliminate them merely to obtain a $pT$-based formulation of the linearized system.
While such an elimination may be tempting when using iterative solvers and pressure-based linear preconditioners such as CPR \cite{roy2020}, care must be taken because the inverse thermodynamic mappings generally exhibit much poorer conditioning across phase transitions than the forward mappings.
For example, at fixed $p$, the mapping $\specEnthalpy(T)$ becomes singular at the phase boundary, whereas the inverse relation $T(\specEnthalpy)$ remains bounded, with $T$ remaining constant throughout the two-phase region.
Analogously, at fixed $T$, the map $p(\specVolume)$ remains well defined across the two-phase region, while the inverse relation $\specVolume(p)$ becomes non-unique or ill-conditioned.
Consequently, the definition of the primary variable $\primvars_{\espec}$ should be preserved consistently throughout all levels of the numerical solution strategy.

\subsection{Isochoric equilibrium as a nonlinear preconditioner}\label{subsec:isochoric-preconditiong}
For the examples in \Cref{sec:examples}, we prescribe time-discontinuous increases in fracture aperture as an idealized representation of rupture-induced dilation.
Laboratory measurements show that such dilation can occur sufficiently rapidly to reduce local fault pressure from several tens of megapascals to vapor pressure, while regions away from the fault respond more gradually \citep{brantut2020}.
The discontinuity reflects the separation between the short time scale of rupture and slip and the much longer time scales of fluid flow and heat transport in porous media.
On the rupture time scale, the fracture pore volume may therefore change before significant mass or energy is transported through the surrounding medium.
Although the examples focus on fracture opening, the numerical treatment introduced below depends only on abrupt changes in pore volume and is not specific to aperture variations.
The same principle could therefore be applied to rapid dilation or compaction of the porous matrix in a fully coupled flow–geomechanics model.

Partitioned flow–geomechanics schemes provide a natural setting for treating such changes.
In fixed-stress splitting, for example, the mechanics and flow subproblems are solved sequentially while the mean total stress is held fixed during the flow solve.
The mechanics update modifies the pore volume, which subsequently enters the flow subproblem as part of the coupled state.
Motivated by this structure, we regard the updated pore volume as the coupling quantity supplied to the flow solver and seek to resolve its immediate thermodynamic consequences before advancing the slower flow and transport processes.

An analogous separation of time scales is already employed in the nonlinear solution procedure described in \Cref{subsec:num-method}: local thermodynamic equilibrium is computed before performing the Schur complement reduction, thereby resolving the fastest physical process.
We extend this idea by applying a nonlinear preconditioning step whenever a sufficiently large pore-volume change is detected at the beginning of a time step.
The updated pore volume is first used to compute a preliminary fluid specific volume while keeping the local fluid mass and fluid internal energy unchanged.
Previous studies have represented the immediate thermodynamic response through either isothermal piston models \cite{weatherley2013} or a separate isenthalpic calculation performed before the post-seismic transport simulation \cite{alfaro2024}.
Here, the rapid event is instead modeled as free expansion of a fixed local fluid amount.
The preliminary fluid specific volume and the pre-change specific internal energy are used as target states for $\specIntEnergy\specVolume$-specified equilibrium calculation.
If isothermal conditions are assumed, $\specVolume T$-specifications are employed instead.
The equilibrium calculation is therefore isochoric at the updated volume, although the preceding mechanical event changes the pore volume.
The resulting thermodynamic state is used to initialize the subsequent coupled nonlinear solve.

The two assumptions represent different idealizations of the short mechanical event: the present work adopts free expansion of a closed local fluid amount, whereas \citet{alfaro2024} model the co-seismic transition as isenthalpic.

As a proof of concept, we restrict our attention to fracture opening.
Let the fracture aperture be described by the time-dependent function $a(t) = a_r \gamma(t)$, where $a_r$ denotes the residual aperture and $\gamma$ a prescribed scaling function.
Let $t_n$ denote the previous converged time level and $t_{n+1} = t_n + \Delta t$ the subsequent time step.
Before entering the global nonlinear iteration, we evaluate the cell-wise pore-volume ratio
\begin{equation}
    \gamma = \frac{(c_{v} \phi a)(t_{n+1})}{(c_{v} \phi a)(t_n)},
\end{equation}
where $c_{v}$ denotes the geometric volume of cells.
If $\gamma \geq \epsilon_{\gamma} > 1$ for a prescribed tolerance $\epsilon_{\gamma}$, we compute the preliminary fluid specific volume
\begin{equation}
    \specVolume_{\star} = \gamma \specVolume(t_n),
\end{equation}
representing the fluid expansion associated with the sudden increase in available pore volume.

For isothermal simulations, equilibrium calculations are performed using the updated specific volume $\specVolume_{\star}$ together with the prescribed temperature $T$.
For thermal simulations, the fluid specific internal energy at the previous time level, $\specIntEnergy(t_n)$, is additionally supplied, and a $\specIntEnergy\specVolume$-equilibrium calculation is carried out.
These specifications are independent of the global problem formulation $\espec$.
The resulting equilibrium state is then used to update all thermodynamic degrees of freedom, thereby providing an improved initial guess for the subsequent global nonlinear iterations.
When specific volume is an independent primary variable of the formulation, it is updated to $\specVolume_{\star}$ as well, ensuring a thermodynamically consistent representation of the imposed pore-volume change.

The proposed strategy should therefore not be regarded as an ad hoc initialization or an artificial numerical correction, but rather as a physics-based nonlinear preconditioning step.
Instead of introducing a purely numerical modification, it models the thermodynamic path by which the fluid reaches the new equilibrium state after a sudden pore-volume increase.
It augments the local equilibrium formulation with an intermediate free-expansion step, after which the governing equations continue to enforce equilibrium with respect to the selected specification $\espec$.
This transition occurs on the same instantaneous time scale as the local equilibrium assumption itself, but is not uniquely prescribed by the governing equations or the equilibrium formulation.
Consequently, the global nonlinear solver is initialized from a state that is both thermodynamically consistent and substantially closer to the solution of the new time step.
The complete nonlinear preconditioning procedure is contained in \Cref{alg:global}.

\begin{center}
\begin{algorithmic}[1]
\Require $T_{end}, \Delta t_{lim}, N, m, \epsilon_a, \epsilon_r, \epsilon_{loc}, \epsilon_{\gamma}, \espec_{glob}, M=\{grid~:\espec_{loc}\}, \Delta t, p(0), T(0), \ofrac(0)$
\State discretize grid
\State initial equilibrium: $\primvars_{\espec_{glob}}(0), \secvars_{\espec_{glob}}(0) \gets \left(p(0), T(0), \ofrac(0)\right)$
\Comment{$pT$-based since initially liquid-saturated}
\State $t_n \gets 0$, $\primvars_{\espec_{glob}}^k\gets\primvars_{\espec_{glob}}(0)$, $\secvars_{\espec_{glob}}^k\gets\secvars_{\espec_{glob}}(0)$
\If{isothermal}
\State $\espec_{pre}\gets \specVolume T$
\Else
\State $\espec_{pre}\gets \specIntEnergy\specVolume$
\EndIf
\State discretize MPFA
\While{$t_n < T_{end}$}
\State update boundary values at $t_n + \Delta t$
\For{$fracture$ in $domains$}
\State $\gamma\gets a(t_n + \Delta t) / a(t_n)$
\If{$\gamma > \epsilon_{\gamma}$}
\State $\secvars_{\espec_{glob}}^k\hookleftarrow$ \Call{NPIPM}{$\espec_{pre},\epsilon_a,\primvars_{\espec_{glob}}^k$, $\secvars^0=\secvars_{\espec_{glob}}^k$} \Comment{on $fracture$ only}
\EndIf
\EndFor
\State $success \gets \texttt{False}$
\For{$k\gets 1,\dots,N$}
\If{$i \mod{m == 0}$}
\State $\Hat{\secvars}\gets 0$
\For{$grid$ in $domains$}
\If{$grid$ in $M$}
\State $\espec_{loc}\gets M[grid]$
\Else
\State $\espec_{loc} \gets \espec_{glob}$
\EndIf
\State $\Hat{\secvars} \hookleftarrow$ \Call{NPIPM}{$\espec_{loc},\epsilon_{loc},\primvars_{\espec_{glob}}^k + \Delta\primvars$, $\secvars^0=\secvars_{\espec_{glob}}^k + \Delta \secvars$} \Comment{on $grid$ only}
\EndFor
\State $\secvars_{\espec_{glob}}^k \gets \Hat{\secvars}$
\EndIf
\State assemble coupled system \eqref{equ:coupled-system-discretized}
\State linear right-preconditioning
\State Schur complement reduction \eqref{equ:coupled-system-reduced}
\State $\Delta \primvars\gets$ solve linear system
\State $\Delta\secvars\gets$ expand Schur complement solution
\State rescale update to physical dimensions
\State $\Delta\primvars\gets$ chop primary thermodynamic state update \Comment{e.g., max $\pm~10\%$ change}
\If{\texttt{phase change detected}}
\State $\Hat{\secvars} \hookleftarrow$ \Call{NPIPM}{$\espec_{glob},\epsilon_{loc},\primvars_{\espec_{glob}}^k + \Delta\primvars$, $\secvars^0=\secvars_{\espec_{glob}}^k + \Delta \secvars$}
\State $\Delta\secvars \gets \Hat{\secvars} - \secvars_{\espec_{glob}}^k$
\EndIf
\State $\delta\gets$ \Call{TRUST REGION}{$\Delta \primvars, \Delta \secvars$}
\State $\primvars_{\espec_{glob}}^k\gets\primvars_{\espec_{glob}}^k + \delta \Delta \primvars$
\State $\secvars_{\espec_{glob}}^k \gets \secvars_{\espec_{glob}}^k + \delta \Delta \secvars$
\State $\residual[glob] \gets$ assemble global residual
\If{ \texttt{converged}}
\State $success \gets \texttt{True}$
\State $\texttt{Break}$
\EndIf
\EndFor
\If{$success$}
\State $t_n \gets t_n + \Delta t$
\State $\Delta t \gets c \Delta t$ \Comment{relax/restrict time step size based on iterations}
\Else
\State $\Delta t \gets 0.5 \Delta t$ \Comment{time step halving}
\If{$\Delta t < \Delta t_{lim}$}
\State $\texttt{Simulation Failure}$
\EndIf
\EndIf
\EndWhile
\end{algorithmic}
\captionsetup{type=algorithm}
\captionof{algorithm}{The global solution strategy.}
\label{alg:global}
\end{center}

\FloatBarrier
\section{Examples}\label{sec:examples}
\begin{figure}[H]
    \centering
    \vspace{-10pt}
    \begin{tikzpicture}
    \draw[color=white, pattern={Lines[angle=45,distance=0.2cm]}, pattern color=black] (3.3,0) rectangle (6.6,2);

    \node[rectangle, thick, draw, minimum width=10cm, minimum height=2cm] (domain) at (5,1) {};
    \node[circle, draw, minimum size=1pt,scale=0.5, fill=black] (injection) at (1.5,1) {};
    \node[circle, draw, minimum size=1pt,scale=0.5, fill=black] (production) at (8.5,1) {};
    \draw[black, thick] (2.5,1) -- (7.5,1);

    \node[yshift=0.2cm] at (injection.north) {\small Injection};
    \node[yshift=-0.25cm] at (injection.south) {\small at $r=r_i,~T=T_i$};
    \node[yshift=0.2cm] at (production.north) {\small Production};
    \node[yshift=-0.25cm] at (production.south) {\small at $p=p_p$};

    \node[rectangle, draw, fill=white, color=white, text=black, minimum width=1cm, minimum height=0.5cm] (permbox)  at (5,1.5) {\small $\absperm_L$};
    \node at (7,1.2) {\small $\Omega_1$};
    \node at (0.3,1.7) {\small $\Omega_0$};
    \end{tikzpicture}
    \caption{Rectangular domain $\Omega_0$ with two point grids serving as injector and producer.
    A single line fracture $\Omega_1$ passes the low-permeability area $\absperm_L$ in the middle.
    Injection rate $r_i$, injection temperature $T_i$ and production pressure $p_p$ are prescribed constants.
    }
    \label{figure:simulation-setup}
\end{figure}
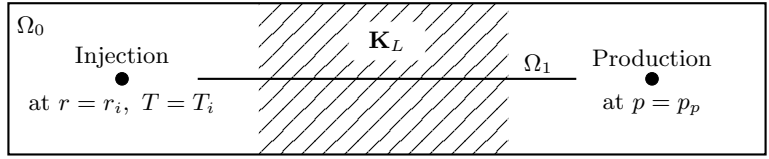

\begin{wraptable}{r}{0.45\textwidth}
\centering
\begin{tabular}{l l}
\toprule
domain & $100$ \unit{\meter} $\times$ $20$ \unit{\meter} \\
$h_{mesh}$ & 1--5 \unit{\meter} \\
$\absperm_{\Omega_0}$ & \num{1e-12} \unit{\meter\squared} \\
$\absperm_l$ & \num{1e-14} \unit{\meter\squared} \\
$\absperm_{\Omega_1}$ & \num{1e-10} \unit{\meter\squared} \\
$\Phi_{\Omega_0}$ & 0.1 \\
$\Phi_{\Omega_1}$ & 1 \\
$\rho_s$ & 2950 \unit{\kilo\gram\per\cubic\meter}\\
$h_s(T)$ & $c_s(T - T_i)$ \unit{\joule\per\kilo\gram} \\
$c_s$ & 603 \unit{\joule\per\kilo\gram\per\kelvin} \\
$\kappa_s$ & 1.6736 \unit{\watt\per\meter\per\kelvin} \\
$\mu_\gamma$ & \num{1e-3} \unit{\pascal\second}\\
$\kappa_\gamma$ & 1 \unit{\watt\per\meter\per\kelvin} \\
$r_i$ & $10\times47134$ \unit{\mole\per\cubic\metre\per\hour}\\
$T_i$ & 450 \unit{\kelvin}\\
$p_p$ & 10 \unit{\mega\pascal}\\
$p(t_0)$ & 10 \unit{\mega\pascal}\\
$T(t_0)$ & 450 \unit{\kelvin}\\
$t_{end}$ & 50 \unit{\day} \\
$t_{\star}$ & 25 \unit{\day}\\
$\Delta t_0$ & 1/2 \unit{\day} \\
$\Delta t_{lim}$ & 1 \unit{\second} \\
\bottomrule
\end{tabular}
\captionof{table}{Domain specifications, initial conditions, rock properties and fluid transport properties.}
\label{table:setup}
\end{wraptable}

\begin{figure}
    \centering
    \includegraphics[width=\textwidth]{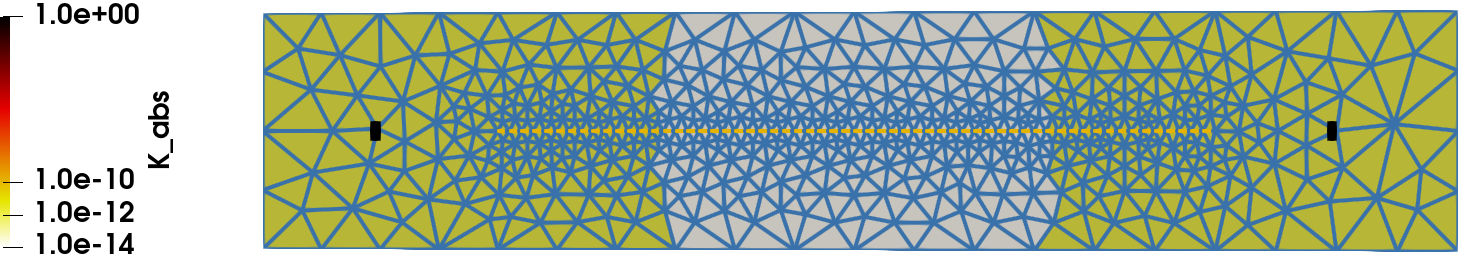}
    \caption{The mesh for the mixed-dimensional domain in \Cref{fig:md-domain}, and the absolute permeability $\absperm$.
    The mesh size decreases gradually from five meters on the external boundary, to one meter on the internal boundary and in the fracture.
    Wells have a permeability of one.
    }
    \label{fig:mesh}
\end{figure}

In this section, we evaluate the proposed volume-based formulation and the isochoric nonlinear preconditioner.
The simulation setup represents a simplified injection and production site, with one injector and one producer.
The rock matrix $\Omega_0$ is assigned a heterogeneous permeability field, where the central region has a lower permeability $\absperm_L$ than the surrounding parts of the domain.
A high-permeability line fracture $\Omega_1$ intersects this region, establishing a preferential flow path between the wells.
We assume that no rock material is present in the fracture and set the fracture porosity to one.
Injection and production conditions are intentionally simplified to study the effects of fracture opening, using a constant and prescribed injection rate, injection temperature, and production pressure.
We model two-phase flow of pure water, also for simplicity.
To isolate the phase change purely induced by mechanical effects, we consider both thermal and isothermal cases.
In the latter case, temperature is a prescribed constant value used for evaluating fluid properties.
The setup is chosen such that the flow is in a steady state before the fracture opening and rapidly returns to a steady state afterward.
The domain is shown in \Cref{figure:simulation-setup}, and the parameters are summarized in \Cref{table:setup}.
The mesh is visualized in \Cref{fig:mesh}.

We compare several equilibrium specifications $\espec$, both with and without the proposed nonlinear preconditioning.
We also conduct a parameter study by introducing a discontinuity in the aperture.
All cases simulate a total of 50 days.
After 25 days, an artificial fracture opening is introduced through a discontinuous increase in the aperture.
More precisely, for $t_{\star} = 25$ \unit{\day} we have
\begin{equation}\label{def:time-aperture}
    a(t) =
    \begin{cases}
        a_r & \text{ if } t < t_{\star},\\
        \gamma a_r & \text{ if } t \geq t_{\star},
    \end{cases} 
\end{equation}
with $\gamma$ being the studied parameter.
Simulation cases are identified using the notation
\begin{center}
    \simcase{(i-)<FLOW>(<PRECOND>)-<$\gamma$>},
\end{center}
where \texttt{FLOW} denotes the equilibrium specification of the governing equations, the optional \texttt{PRECOND} denotes the equilibrium specification used by the nonlinear preconditioner, and $\gamma$ is the fracture aperture scaling factor.
If the optional specification is absent, the preconditioner is not used.
If the optional suffix \texttt{i} is present, the case denotes an isothermal simulation without the energy balance equation. 
For example, \simcase{i-pT(\specVolume T)-1.5} denotes an isothermal simulation using the $pT$-formulation together with $\specVolume T$-equilibrium calculations in the nonlinear preconditioner, where the fracture aperture is increased by a factor of 1.5.

Throughout this section, the last converged time step preceding the aperture increase is denoted by $t_{-1}$.
In particular, the setup is such that the flow is in a steady state for $t \leq t_{-1}$ and reaches the steady state after some time $t > t_{\star}$.
We denote transient times with the symbol $\tau$ and distinguish between the gas transient $\tau_g$ and the pressure transient $\tau_p$.
The gas transient is measured using the gas content in the fracture $\sat[\text{frac}]$, which is greater than zero during that period.
The pressure transient is measured using the $L_2$-norm and the steady-state solution for $p$.
Once the $L_2$-difference is sufficiently small, the transient period is deemed over.

\subsection{Volume-based models}\label{subsec:volume-examples}
\begin{figure}
    \centering
    \begin{subfigure}{0.85\textwidth}
        \includegraphics[width=\linewidth]{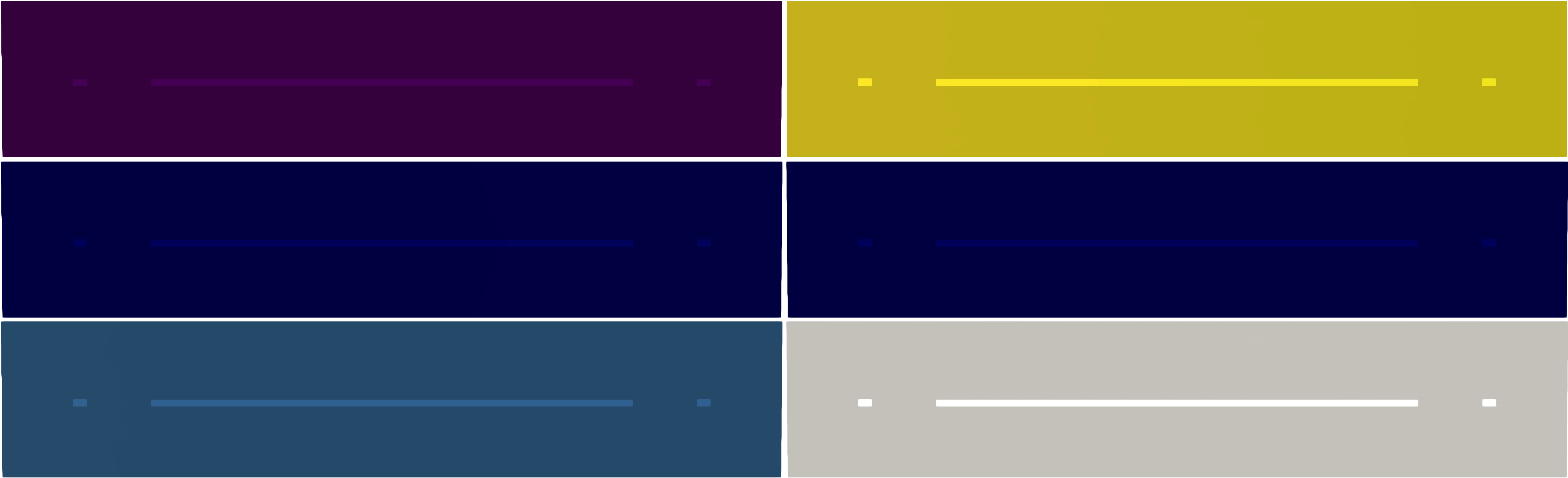}
        \caption{}
        \label{fig:simulation-iso-a}
    \end{subfigure}
    \begin{subfigure}{0.85\textwidth}
        \includegraphics[width=\linewidth]{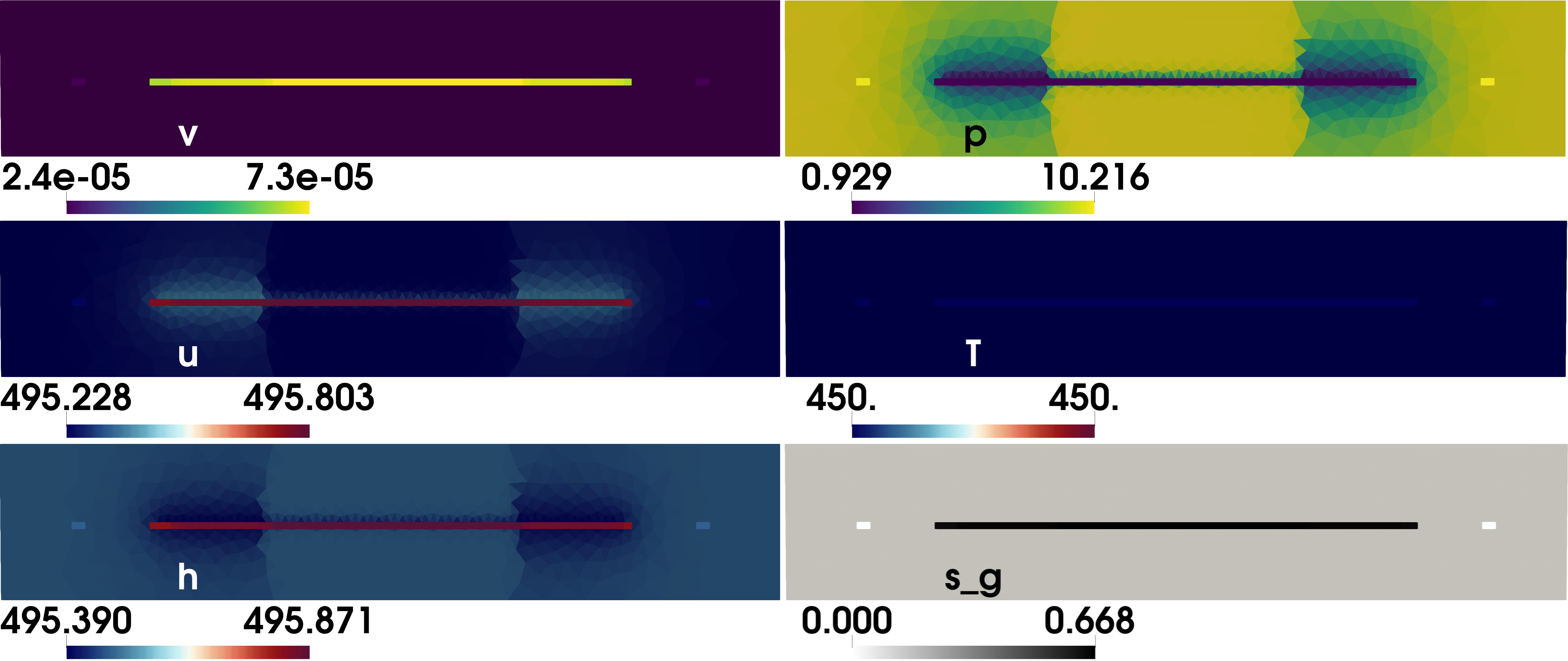}
        \caption{}
        \label{fig:simulation-iso-b}
    \end{subfigure}
    \begin{subfigure}{0.85\textwidth}
        \includegraphics[width=\linewidth]{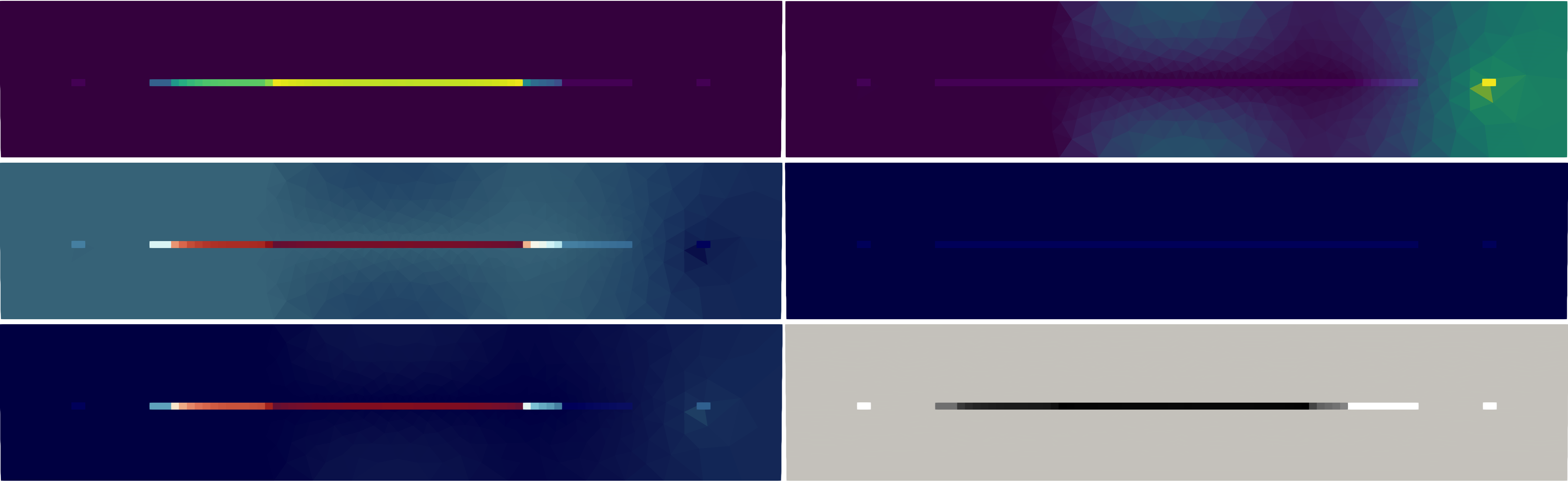}
        \caption{}
        \label{fig:simulation-iso-c}
    \end{subfigure}
    \begin{subfigure}{0.85\textwidth}
        \includegraphics[width=\linewidth]{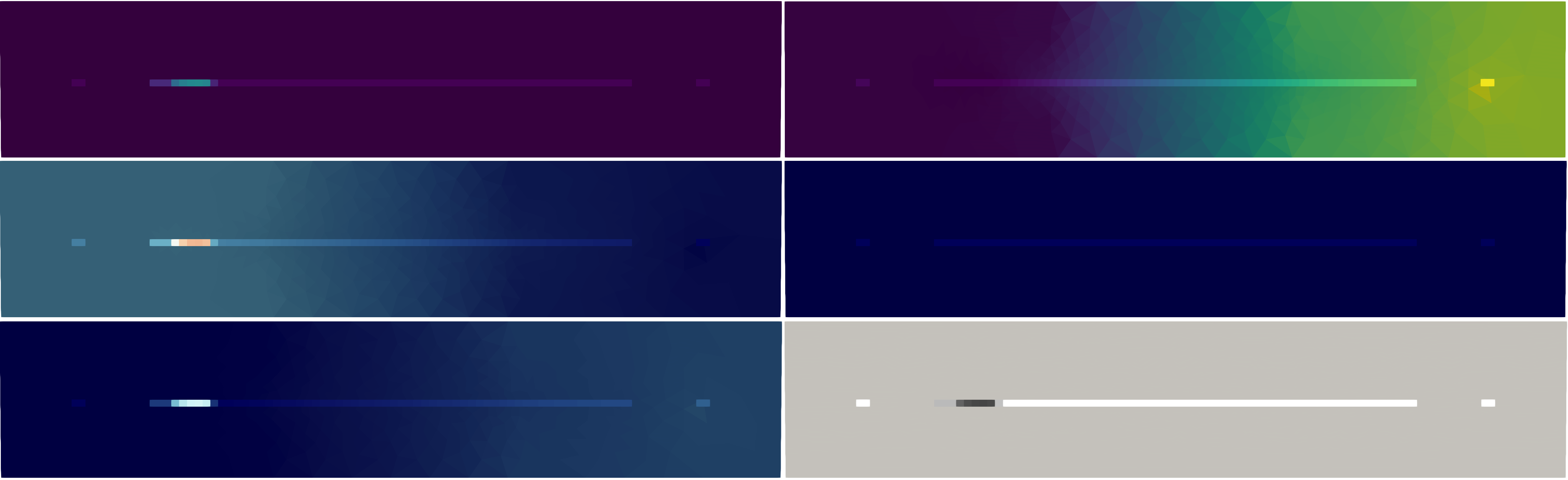}
        \caption{}
        \label{fig:simulation-iso-d}
    \end{subfigure}
    \vspace{-10pt}
    \caption{Simulation \simcase{i-vT(vT)-3.0}. Appearance of gas upon fracture opening. 
    Left to right, top to bottom: $\specVolume$ \myunit{\mol\per\cubic\meter}, $p$ \myunit{\mega\pascal}, $\specIntEnergy$ \myunit{\kilo\joule}, $T$ \myunit{\kelvin}, $\specEnthalpy$ \myunit{\kilo\joule}, $\sat[g]$ [-].
    (a) shows steady state before fracture opening one second before $t_{\star}$.
    (b) shows the gas formation at $t_{\star}=25$ days.
    (c) shows gradual disappearance of gas phase after five minutes.
    (d) shows $p$ re-entering steady state and an almost fully recompressed fluid again after one hour.
    }
    \label{fig:simulation-iso}
    \vspace{-20pt}
\end{figure}

First, we present the simulation results of simulation case \simcase{i-vT(vT)-3.0} in order to provide a visual impression of the process.
\Cref{fig:simulation-iso} shows snapshots in time before, during, and after the fracture opening.

\begin{figure}
    \centering
    \begin{subfigure}{0.85\textwidth}
        \includegraphics[width=\linewidth]{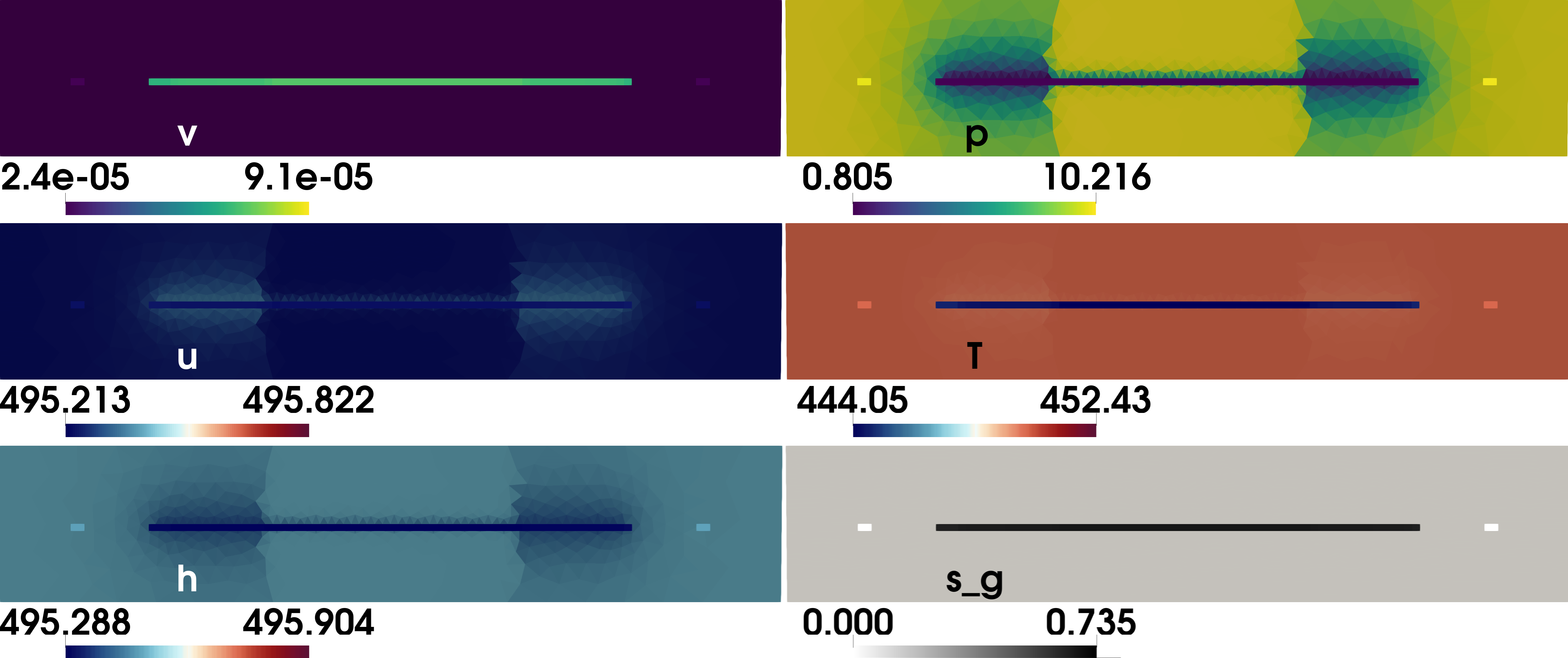}
        \caption{}
        \label{fig:simulation-thermal-a}
    \end{subfigure}
    \begin{subfigure}{0.85\textwidth}
        \includegraphics[width=\linewidth]{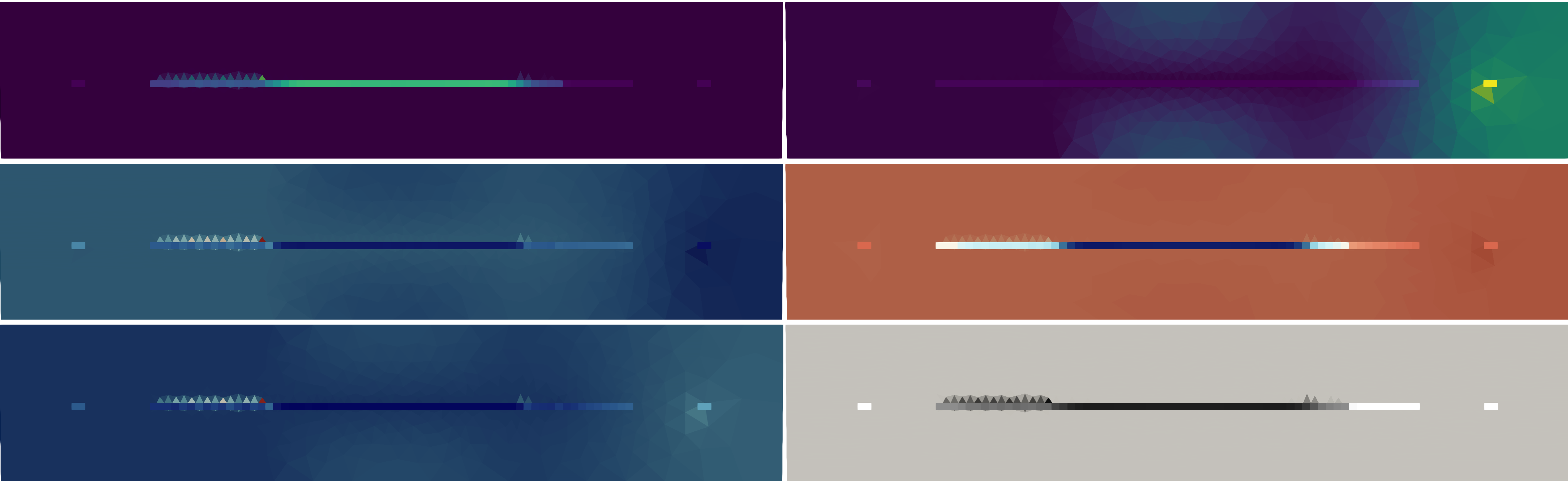}
        \caption{}
        \label{fig:simulation-thermal-b}
    \end{subfigure}
    \begin{subfigure}{0.85\textwidth}
        \includegraphics[width=\linewidth]{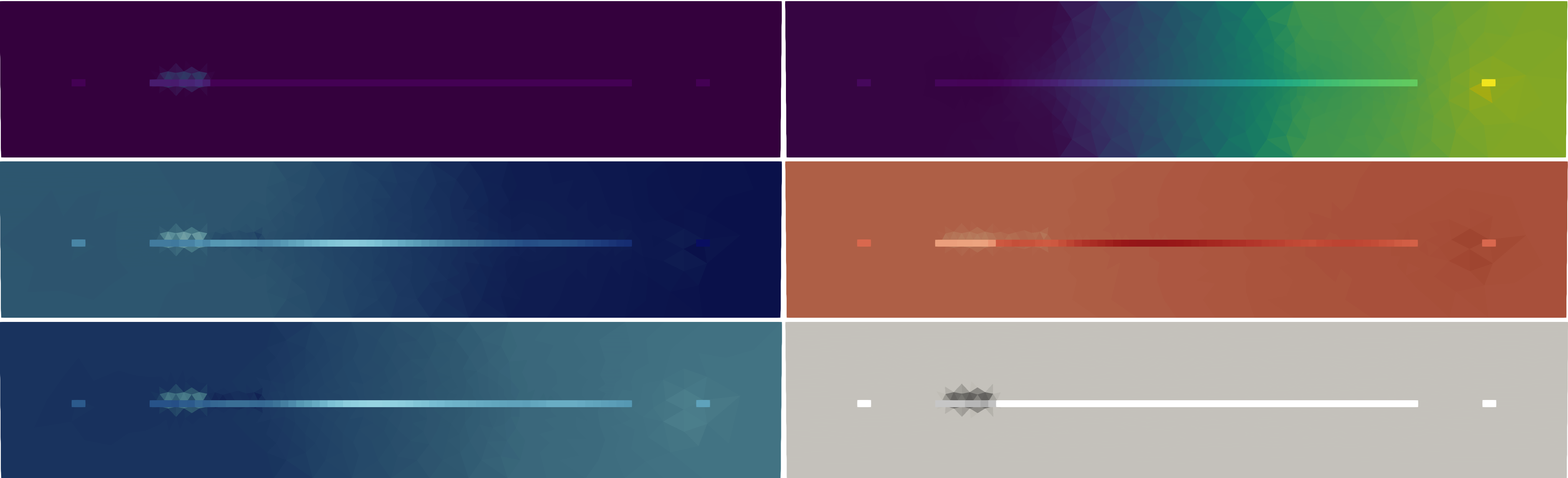}
        \caption{}
        \label{fig:simulation-thermal-c}
    \end{subfigure}
    \begin{subfigure}{0.85\textwidth}
        \includegraphics[width=\linewidth]{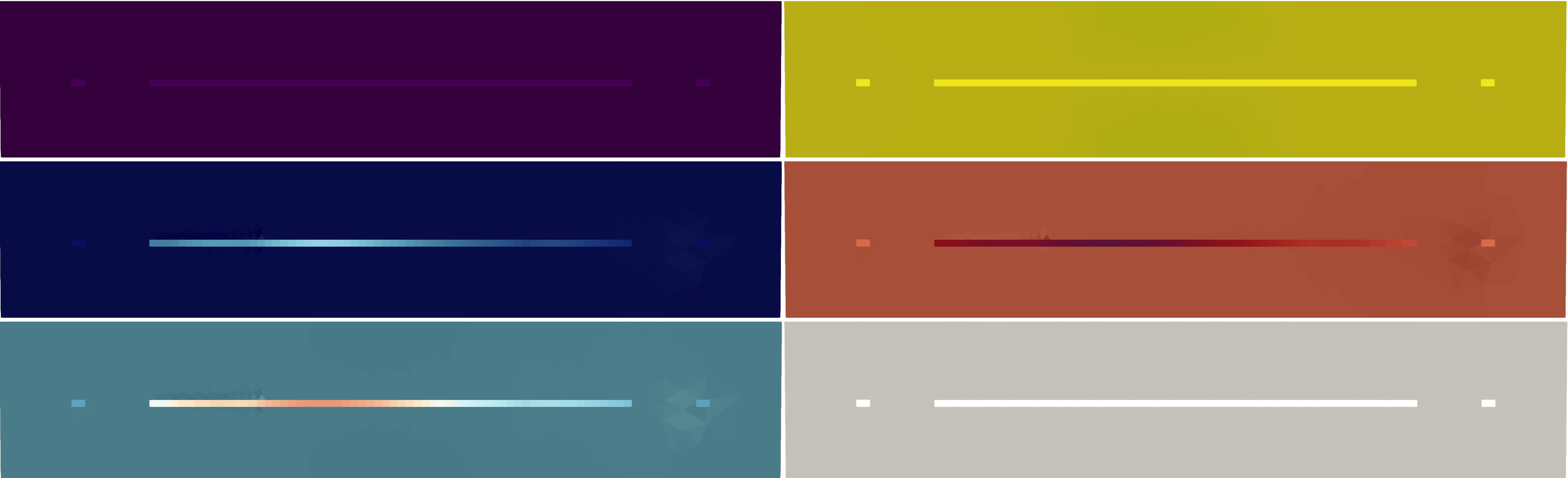}
        \caption{}
        \label{fig:simulation-thermal-d}
    \end{subfigure}
    \vspace{-10pt}
    \caption{Thermal simulation \simcase{vT(uv)-3.0}.
    (a) shows shows the fluid response at $t_{\star}$, now including a drop in $T$.
    (b) shows gradual recompression and an energetic response due to the thermal gradient between fracture and matrix after five minutes.
    (c) shows traces of gas after one hour and an increase of fluid internal energy due to conduction.
    (d) shows a re-compressed, near-steady-state fluid and increased internal energy in the fracture after 90 minutes.
    }
    \label{fig:simulation-thermal}
    \vspace{-20pt}
\end{figure}

\Cref{fig:simulation-iso-a} shows the solution in steady state at the last converged time step, $t_{-1}$, before the aperture jump.
The simulation is isothermal, and the entire domain is initially liquid saturated.
Pressure ranges from approximately 10.2 \unit{\mega\pascal} at the injector to the prescribed 10 \unit{\mega\pascal} at the producer.

\Cref{fig:simulation-iso-b} illustrates the immediate response to the fracture opening.
The sudden increase in pore volume causes an instantaneous decompression of the fluid, with the pressure dropping by more than 9 \unit{\mega\pascal} inside the fracture.
The system subsequently enters the transient phase.
The heterogeneous permeability field strongly influences the propagation of this pressure disturbance.
While the decompression originates within the fracture, the surrounding low-permeability region delays its propagation, whereas the high-permeability regions exhibit a substantially stronger initial response.

We observe the appearance of gaseous water.
Only inside the fracture does the thermodynamic state cross the liquid--vapor phase boundary.
Gas nucleation is therefore confined to the fracture.
The gas saturation is largest in the central part of the fracture, reaching approximately 67\%, while the fracture tips exhibit lower gas saturations.

In the isothermal simulations, temperature is prescribed and therefore remains constant throughout the process.
Consequently, both the specific internal energy and the specific enthalpy exhibit only minor variations.
For liquids, the internal energy depends primarily on temperature and only weakly on pressure, explaining the small changes observed in $\specIntEnergy$.
Likewise, the specific enthalpy remains nearly constant.
Although the pressure work $p\specVolume$ changes considerably during the transient, the decrease in pressure largely compensates for the increase in specific volume, resulting in only small changes of the total enthalpy.

\Cref{fig:simulation-iso-c} shows the solution after five minutes.
The low-pressure region has propagated into the low-permeability section of the reservoir, while a pronounced decompression develops around the injection well.
Pressure recovery begins at the production side of the domain, where pressure is prescribed, whereas the injector imposes a constant mass flux and therefore cannot immediately restore the pressure.
The observed profile is hence a direct consequence of the chosen boundary conditions, where we fixed $p$ in the right point grid.

As fluid continues to enter the domain, the expanded pore volume is progressively replenished, and the fluid undergoes recompression.
Gas disappears first near the production side of the fracture, where pressure recovery begins, and subsequently vanishes throughout the remaining fracture.
Although injection continues throughout the transient, the injected mass is initially consumed by filling the newly created pore volume, resulting in only negligible advective transport over this short time interval.

The specific volume gradually returns to values characteristic of the liquid phase, accompanied by a corresponding recovery of the energetic state variables.

\Cref{fig:simulation-iso-d} shows the solution one hour after the fracture opening.
The gas phase has nearly disappeared, and the pressure field has largely recovered.
The strong pressure contrast between the low- and high-permeability regions is no longer visible, and the solution gradually approaches the new steady state.
Approximately 1.7 hours after the aperture jump, the transient has fully decayed, and the simulation has returned to steady-state conditions that are identical to those displayed in \Cref{fig:simulation-iso-a}.

Given the short duration of the transient, one might initially expect the fluid to return to the same thermodynamic state as in the steady-state solution preceding fracture opening.
The thermal simulation shows, however, that this is not the case.
Although a period of several hours is short on a reservoir time scale, it is sufficient for appreciable thermal interaction between the fluid-filled fracture and the surrounding hot rock matrix.
The results for simulation case \simcase{vT(uv)-3.0} are shown in \Cref{fig:simulation-thermal}, now including the energy balance and the associated thermodynamic state variables.
As in the isothermal case, the fluid is initially at 450 \unit{\kelvin}, and the injection temperature is fixed at this value for demonstration purposes.
The nonlinear preconditioning step instead employs the $\specIntEnergy\specVolume$-specification, consistent with the assumption of constant specific internal energy during free expansion.

\Cref{fig:simulation-thermal-a} shows an initial response similar to that of the isothermal simulation in \Cref{fig:simulation-iso-b}.
Gas forms in the fracture, accompanied by a pronounced decrease in pressure within the fracture and its immediate surroundings.
In the thermal simulation, the free-expansion step additionally produces cooling of the fluid for the thermodynamic state considered here.
The temperature decreases by approximately 6 \unit{\kelvin}.
Compared with the isothermal case, the pressure drop is approximately 0.12 \unit{\mega\pascal} larger, and the maximum gas saturation increases to 73.5\%.
The lower pressure is accompanied by an increase in $\specVolume$, while the changes in $\specIntEnergy$ and $\specEnthalpy$ reflect the thermal and mechanical evolution occurring during the converged time step.

\Cref{fig:simulation-thermal-b} shows the state after five minutes.
As in \Cref{fig:simulation-iso-c}, the pressure disturbance has propagated through the domain, with the low-permeability region delaying its advance.
The cooling of the fracture establishes a temperature gradient between the fracture and the surrounding matrix.
Since the total energy of the coupled fluid--rock system is conserved, thermal conduction redistributes energy from the hotter rock toward the cooled fracture.
Under the assumption of local thermal equilibrium, the rock and all fluid phases share a common temperature, so this energy transfer is reflected in an increase in the fluid's specific internal energy.
Unlike in the isothermal simulation, gas is therefore no longer confined to the fracture but also appears in the decompressed region to its right.
This additional phase change results from the combined evolution of pressure, specific volume, and internal energy rather than from decompression alone.

After one hour, gaseous water remains only near the left fracture tip toward the injector, as shown in \Cref{fig:simulation-thermal-c}.
By this time, conductive energy transfer from the rock has increased the internal energy of the fluid in and around the fracture.
The fluid, therefore, undergoes recompression from a thermodynamic state that differs from the state reached in the isothermal simulation.
After pressure recovery, shown in \Cref{fig:simulation-thermal-d}, the temperature in the fracture is approximately 452 \unit{\kelvin}, compared with 450 \unit{\kelvin} before fracture opening.
Besides the slightly larger initial pressure drop and maximum gas saturation, this residual temperature increase constitutes the principal difference from the isothermal case (see steady-state solution in \Cref{fig:simulation-iso-a}).
Once the expanded pore volume has been replenished and advective transport becomes appreciable again, the warmer fluid is gradually transported toward the producer.

The thermal simulation shows that the response to mechanically induced phase change is not limited to pressure and phase saturation.
For the present state, free expansion initially cools the fluid and establishes a temperature gradient between the fracture and the surrounding rock.
Heat conduction subsequently transfers energy toward the cooled region, so that the fluid contains more internal energy when recompression occurs and reaches a slightly higher final temperature.
The magnitude of this effect is amplified here by the persistent fracture opening and the threefold increase in aperture.
Nevertheless, the results indicate that thermal exchange with the rock matrix can influence both the transient phase behavior and the post-transient fluid state, even when the mechanically induced phase change itself is short-lived.

This sequence is consistent with the post-seismic calculations of \citet{alfaro2024}, in which rapid decompression establishes a reduced initial fracture temperature followed by conductive reheating and, for larger openings, propagation of the thermal disturbance into the surrounding damage zone.
In contrast to models that assume immediate restoration of an isothermal state through heat exchange with the wall rock \cite{weatherley2013}, the present formulation resolves the finite thermal evolution within the coupled energy-transport problem.

\subsection{Comparison with pressure-based formulations}\label{subsec:hybrid}
\begin{figure}
    \centering
    \includegraphics[width=0.65\textwidth]{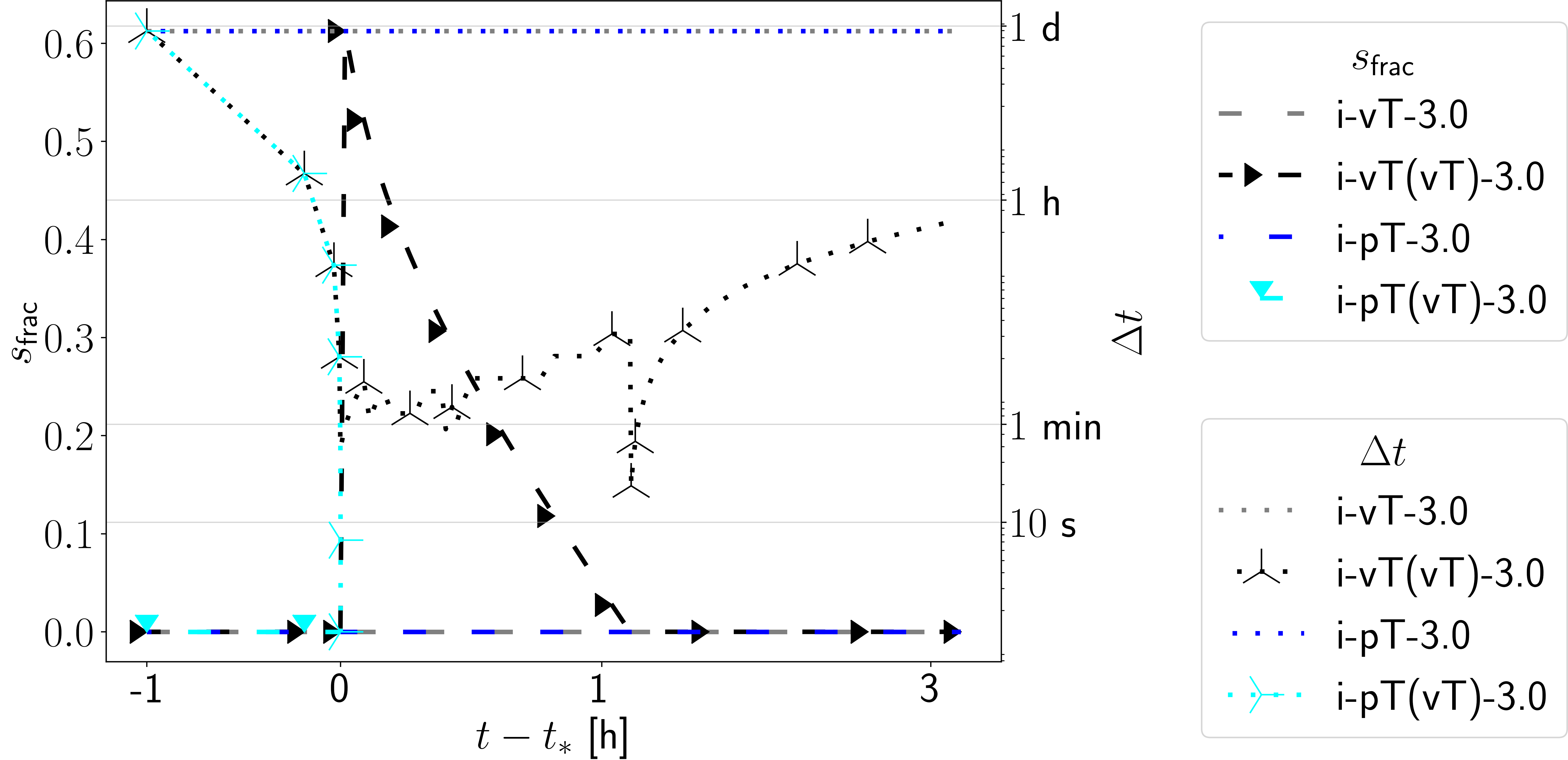}
    \caption{Comparison of time step size $\Delta t$ and volumetric gas content in the fracture $\sat[\text{frac}]$ for the isothermal pressure- and volume-based formulations.
    The volume-based formulation with nonlinear preconditioning (\simcase{i-vT(vT)-3.0}) is the only case which successfully resolves the gas transient.
    }
    \label{fig:dt-p-based-isothermal}
\end{figure}

In this section, we compare the volume-based formulations with their pressure-based counterparts and investigate the effect of the proposed isochoric nonlinear preconditioner.
\Cref{fig:dt-p-based-isothermal} shows the volumetric gas content in the fracture $\sat[\text{frac}]$ and the adaptive time-step size for the equilibrium specifications $\espec\in\{(p, T), (\specVolume, T)\}$, both with and without nonlinear preconditioning.
To emphasize the differences between the formulations, only the most demanding case, $\gamma=3.0$, is considered.

As anticipated, the pressure-based formulation without preconditioning, \simcase{i-pT-3.0}, does not resolve the transient gas phase.
Since $pT$-equilibrium calculations cannot uniquely represent the two-phase region for pure water, the nonlinear solver advances directly from the converged solution at $t_{-1}$ to a liquid-saturated state after the fracture opening.
The adaptive time-stepping strategy, therefore, steps over the transient without detecting it.
Introducing the nonlinear preconditioner, as in \simcase{pT(vT)-3.0}, allows the transient gas phase to be identified during the preconditioning step.
Nevertheless, due to the inadequacy of the $pT$-based formulation for multiphase flow, the preconditioner leads to repeated failures and reductions in time step size until the lower limit for the time step size is reached (one second), and the simulation fails.

Interestingly, the volume-based formulation alone, \simcase{i-vT-3.0}, also fails to resolve the gas phase and steps over the transient period.
This demonstrates that the equilibrium specification alone is insufficient when the transient is shorter than the adaptive time step.
The nonlinear preconditioner, therefore, plays a crucial role in identifying rapid thermodynamic state changes induced by abrupt pore-volume variations.
Among the investigated isothermal setups, only \simcase{i-vT(vT)-3.0}, combining the $\specVolume T$-formulation with $\specVolume T$-based preconditioning, successfully resolves the transient gas phase.
The pronounced pressure drop associated with gas formation nevertheless requires the adaptive time-stepping algorithm to reduce the time-step size to approximately 38 \unit{\second}, indicating that further improvements to the nonlinear solution strategy remain desirable.

\begin{figure}
    \centering
    \includegraphics[width=0.65\textwidth]{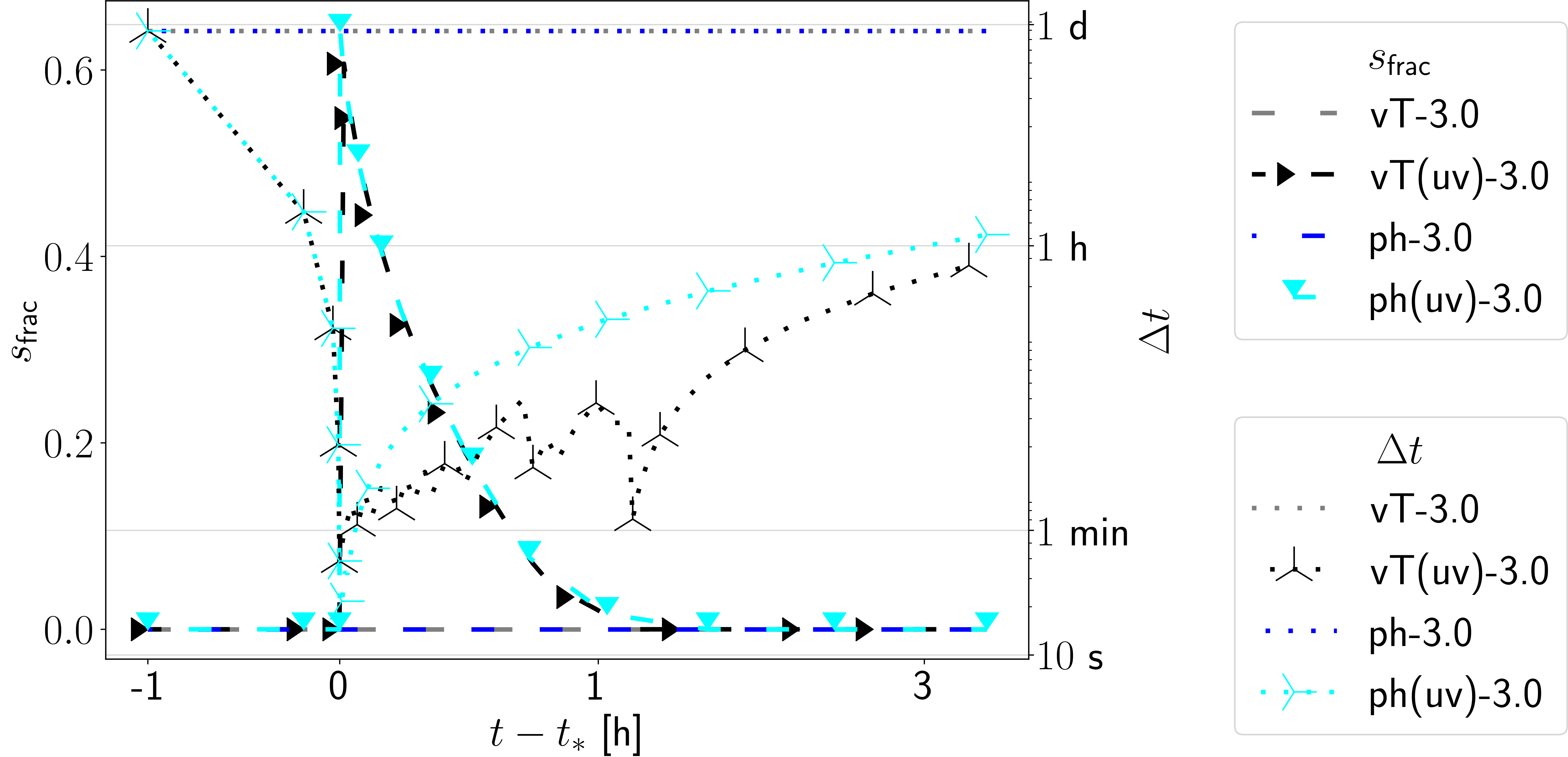}
    \caption{Time step size and volumetric gas content for the thermal formulations.}
    \label{fig:dt-p-based-thermal}
\end{figure}

\Cref{fig:dt-p-based-thermal} presents the corresponding investigation for the thermal simulations.
The global equilibrium specifications are $(p,\specEnthalpy)$ and $(\specVolume,T)$, while the nonlinear preconditioner performs an equilibrium calculation at constant $\specIntEnergy$ and $\specVolume$.
All cases reach the final simulation time.
However, only the preconditioned formulations resolve the transient phase change.
Since the thermal model additionally evolves $\specEnthalpy$, the preconditioned pressure-based formulation also advances successfully to the final simulation time of 50 \unit{\day}, while capturing the transient gas generation.
In contrast, both formulations without nonlinear preconditioning again advance across the transient in a single time step and consequently fail to resolve the temporary appearance of the gas phase.

Compared with \simcase{vT(uv)-3.0}, case \simcase{ph(uv)-3.0} exhibits greater robustness with respect to the adaptive time stepping.
While the $p\specEnthalpy$-formulation increases the time step monotonically after the fracture opening at $t_{\star}$, the volume-based formulation undergoes several time step reductions caused by failed Newton iterations.
Both preconditioned formulations initially reduce the time step to approximately 38 \unit{\second} to advance from $t_{-1}$ to $t_{\star}$.
We attribute the reduced robustness of the volume-based formulation to the behavior of the function $p(\specVolume)$ during recompression.
As the fluid transitions from the two-phase region back into the single-phase liquid region, the pressure becomes highly sensitive to changes in specific volume since liquid water is nearly incompressible.
Consequently, small changes in $\specVolume$ produce comparatively large pressure variations, increasing the nonlinearity of the problem and necessitating smaller time steps to maintain robust Newton convergence.

Finally, we emphasize that almost all cases without preconditioning reach the final simulation time of 50 \unit{\day}.
This should not be interpreted as all formulations resolving the same physical process.
Instead, the discrepancy arises from the short-lived nature of the gas transient relative to the adaptive time-step size.
In the cases \simcase{i-pT-3.0}, \simcase{i-vT-3.0}, \simcase{ph-3.0}, and \simcase{vT-3.0}, the time step spanning $t_{-1}$ and $t_{\star}$ exceeds the transient duration $\tau_g$.
Consequently, the numerical solution advances directly from the pre-expansion state to the post-transient steady state without resolving the intermediate gas phase.
Although this issue could be addressed by manually reducing the time step around the prescribed fracture opening, such an approach relies on prior knowledge of the event.
Instead, the objective of the proposed nonlinear preconditioner is to exploit the underlying physics of rapid pore-volume changes to detect and resolve such transient phenomena automatically.
In this sense, the preconditioner serves not only as a nonlinear acceleration technique but also as a physics-based detector of transient thermodynamic events.

\subsection{Study of rapid changes in aperture}\label{subsec:aperture}
In the final set of examples, we investigate the influence of the aperture scaling factor.
\citet{lee2002} reported measurements of fracture apertures in granite and marble, with mean apertures of 0.69/0.54 \unit{\mm} and maximum apertures of 2.62/1.53 \unit{\mm}, respectively. Motivated by the observed maximum-to-mean aperture ratios, we consider aperture scaling factors of up to three.
All simulations in this section employ scheduled time steps, enforcing $\Delta t_{\star} = t_{\star} - t_{-1} = 1$ \unit{\second}, followed by a fixed time step size of one minute during the transient period to facilitate a consistent comparison between all cases.

\begin{figure}
    \centering
    \includegraphics[width=0.7\linewidth]{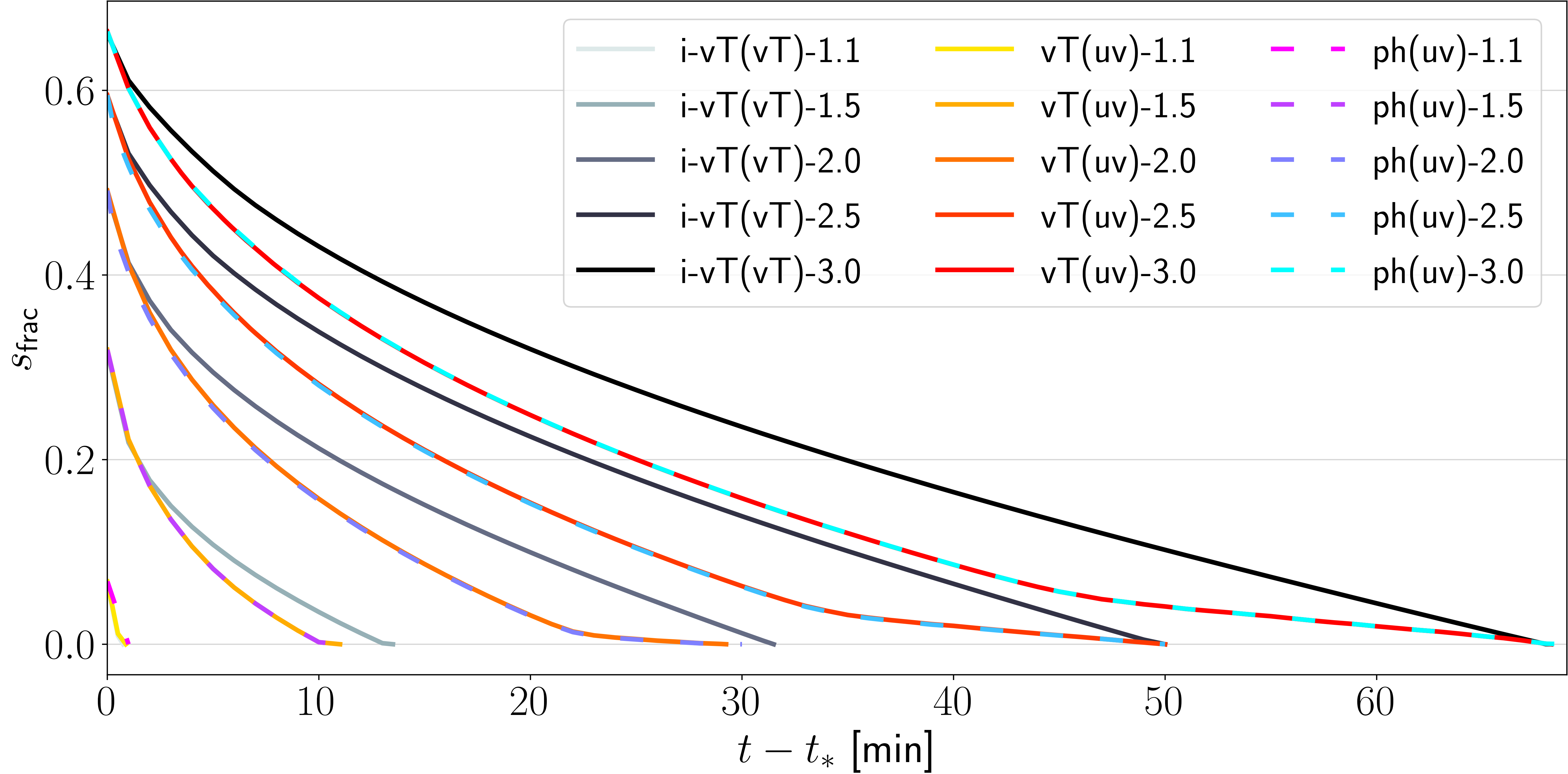}
    \caption{Volumetric gas content in fracture during transient for all successful formulations.}
    \label{fig:gas-content}
\end{figure}

\begin{figure}
    \centering
    \includegraphics[width=0.5\linewidth]{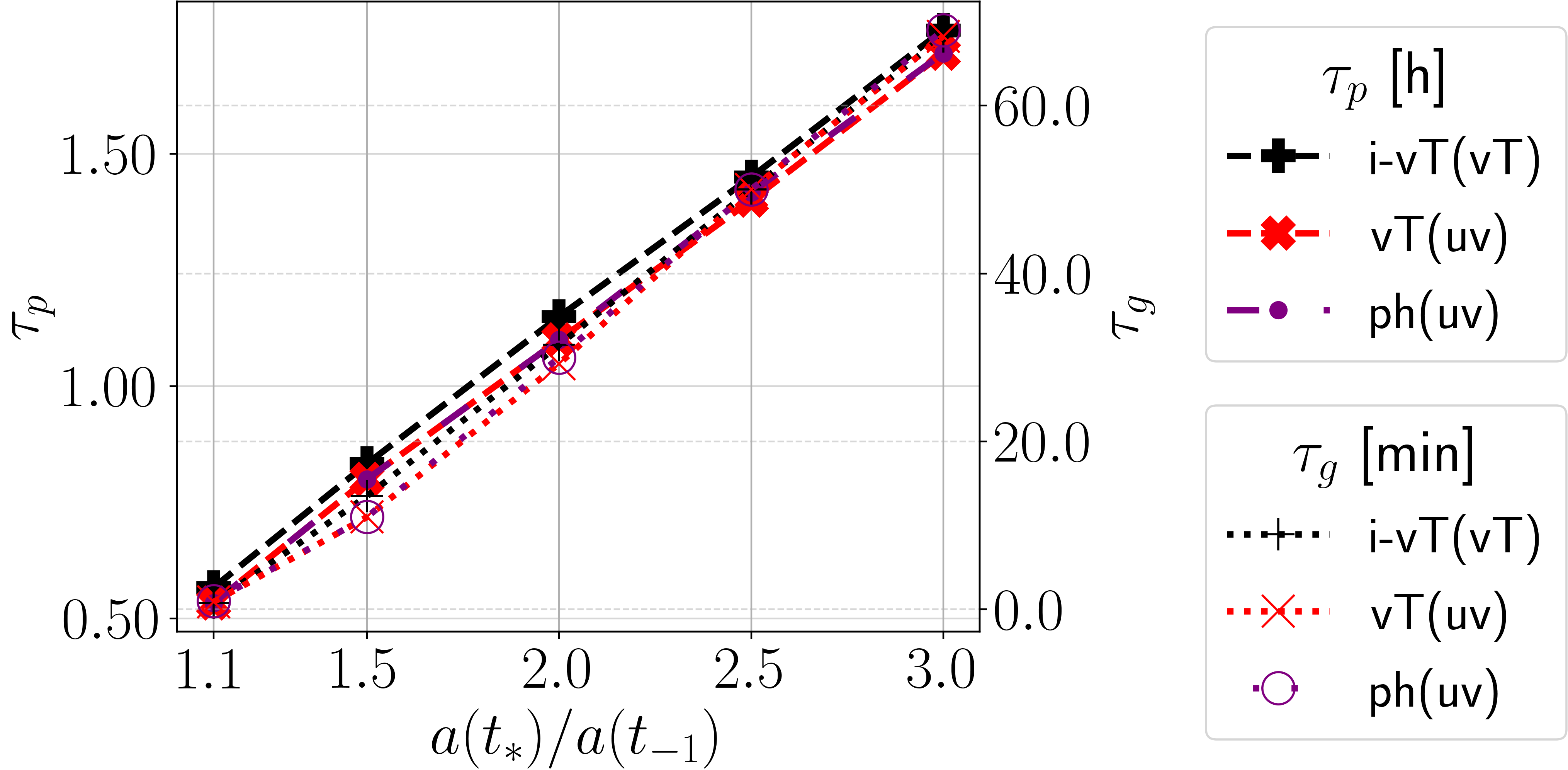}
    \caption{Pressure transient $\tau_p$ and gas transient $\tau_g$ duration per scale factor.}
    \label{fig:transient-duration}
\end{figure}

\Cref{fig:gas-content} shows the volumetric gas content in the fracture, $\sat[\text{frac}]$, as a function of time, while \Cref{fig:transient-duration} summarizes the durations of the pressure and gas transients.
As expected, larger increases in pore volume result in stronger fluid expansion, leading to more pronounced phase separation and a longer-lived gas phase.
The maximum volumetric gas content increases monotonically with the aperture scaling factor for both isothermal and thermal simulations, ranging from 6.51\% for a factor of 1.1 to 66.36\% for a factor of 3.0.
The occurrence of partial rather than complete vaporization is consistent with the calculations of \citet{alfaro2024}, who obtained co-seismic vapor fractions of approximately 10\%–17\% and local post-seismic vapor quantities of the order of \num{1e-1}.
A direct quantitative comparison is not intended, since the thermodynamic initial states, opening geometries, and precise definitions of the reported vapor quantities differ between the two studies.

Although the thermal and isothermal simulations predict nearly identical gas-transient durations, the evolution of the gas content differs noticeably.
Immediately after free expansion, the lower temperature in the thermal simulations favors condensation and produces a steeper initial decline in gas content.
Subsequent conductive heat transfer from the surrounding rock toward the cooled fracture slows condensation, causing the thermal gas-content curves to flatten relative to their isothermal counterparts.

The durations of both the pressure and gas transients increase monotonically with the aperture scaling factor.
The pressure transient ranges from approximately 32 to 106.2 minutes, while the gas transient ranges from 45 seconds to 69 minutes.
The close agreement between the thermal and isothermal durations indicates that recovery is governed primarily by the replenishment of the enlarged pore volume rather than by heat transfer.
Since the injection rates are identical and the temperature variations change the density of liquid water only marginally, thermal effects modify the thermodynamic path without substantially altering the recovery timescale.

Furthermore, the thermal simulations obtained with the volume- and pressure-based formulations produce virtually identical results.
This agreement is expected.
The choice of primary transport variables merely changes the parameterization of the thermodynamic state space, whereas the governing equations describe the same physical trajectory through that space.
Provided both formulations are solved accurately, the resulting physical solution is therefore independent of the chosen equilibrium specification.

The resulting minute-to-hour persistence of the gas phase lies within the broad range reported for dilation-induced transients at different spatial scales.
In the laboratory experiments of \citet{brantut2020}, the near-vapor-pressure state persisted for approximately 250 seconds before pressure recovery became pronounced.
In the simulations of \citet{alfaro2024}, the pressure within the fracture recovered in less than one second, whereas a propagating region containing more than 5\% vapor persisted for up to approximately 2.8 hours.
The comparison indicates that the equilibrium transition itself is effectively instantaneous, while the duration of the observable pressure and phase transients is governed by the surrounding transport problem, including permeability, geometry, and available recharge pathways.

\begin{figure}
    \centering
    \begin{subfigure}{0.55\textwidth}
        \includegraphics[width=\linewidth]{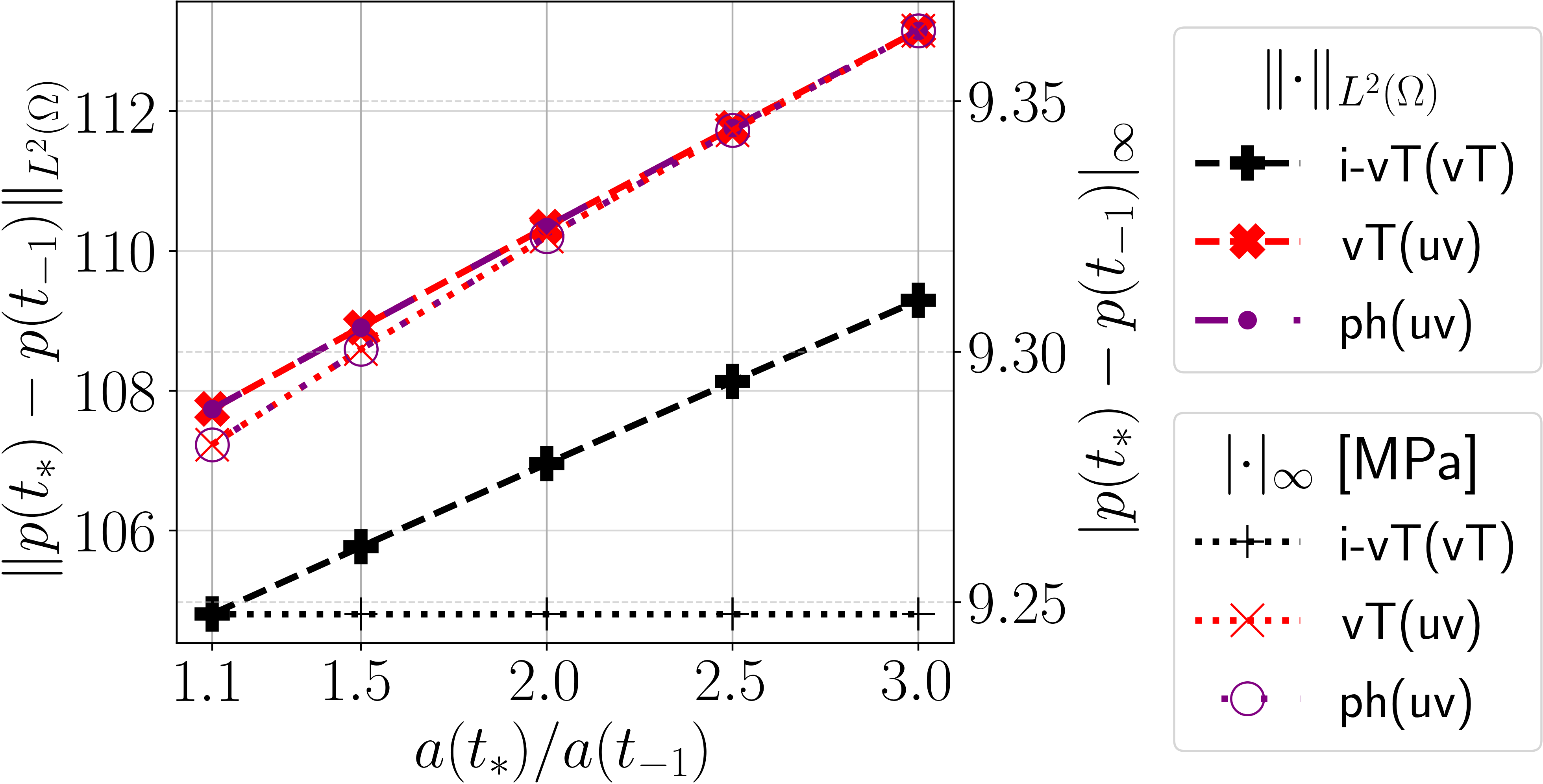}
        \caption{Pressure drop.}
        \label{fig:p-drop-per-a}
    \end{subfigure}
    \hfill
    \begin{subfigure}{0.4\textwidth}
        \includegraphics[width=\linewidth]{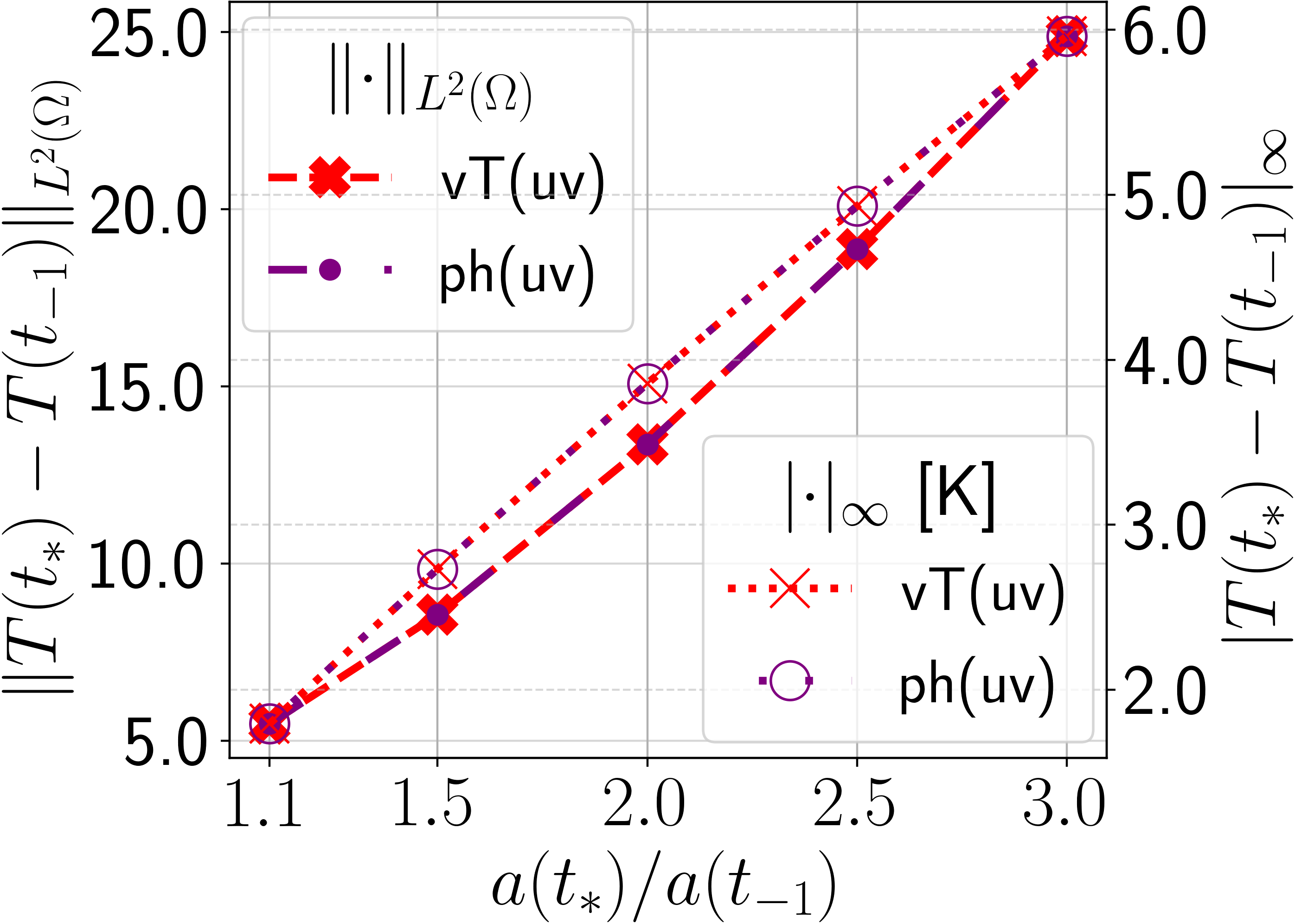}
        \caption{Temperature drop.}
        \label{fig:T-drop-per-a}
    \end{subfigure}
    \caption{Changes in pressure and temperature with increasing aperture scaling factor at time of the fracture opening.}
    \label{fig:drops}
\end{figure}

In \Cref{fig:p-drop-per-a}, we compare the pressure drop immediately after fracture opening.
Two measures are considered: the $L_2$-norm, $\lVert\cdot\rVert_{L_2(\Omega)}$, over the entire mixed-dimensional domain and the point-wise infinity norm, $\lvert\cdot\rvert_{\infty}$.
In both cases, the steady-state solution at $t_{-1}$ immediately preceding fracture opening serves as the reference state.

The first observation is that the pressure response increases monotonically with the aperture scaling factor.
Furthermore, the thermal simulations again produce identical results for the pressure- and volume-based formulations, confirming that both formulations describe the same physical response.

The second observation is that the isothermal simulations exhibit an almost constant maximum pressure drop of approximately 9.25 \unit{\mega\pascal}, independent of the aperture scaling factor.
This behavior is consistent with the phase diagram shown in \Cref{fig:vT-diagram}.
Once the thermodynamic state enters the two-phase region, additional expansion is accommodated primarily through vapor generation rather than by a further reduction in pressure at constant temperature.

The thermal simulations, by contrast, display a gradually increasing maximum pressure drop, ranging from 9.28 \unit{\mega\pascal} to 9.36 \unit{\mega\pascal}.
This behavior follows directly from the increasing temperature drop shown in \Cref{fig:T-drop-per-a}.
The coupled energy balance therefore captures the additional thermodynamic effects associated with free expansion, leading to a slightly lower pressure than predicted by the corresponding isothermal simulations.

Although the present examples are not intended to reproduce a particular faulting event, the magnitude of the pressure reduction is consistent with previous experimental and numerical evidence.
In the high-enthalpy scenario of \citet{alfaro2024}, an initial pressure of 8.69 \unit{\mega\pascal} is reduced below 1 \unit{\mega\pascal} for sufficiently large co-seismic openings.
Moreover, the slip experiments of \citet{brantut2020} produced on-fault pressure reductions ranging from several megapascals to values of approximately 14 \unit{\mega\pascal}, while the main rupture reduced the local pressure from 30 \unit{\mega\pascal} to vapor pressure.
These comparisons support the physical plausibility of substantial dilation-induced depressurization, although the differing scales and boundary conditions preclude direct quantitative validation.

\begin{figure}
    \centering
    \includegraphics[width=0.7\textwidth]{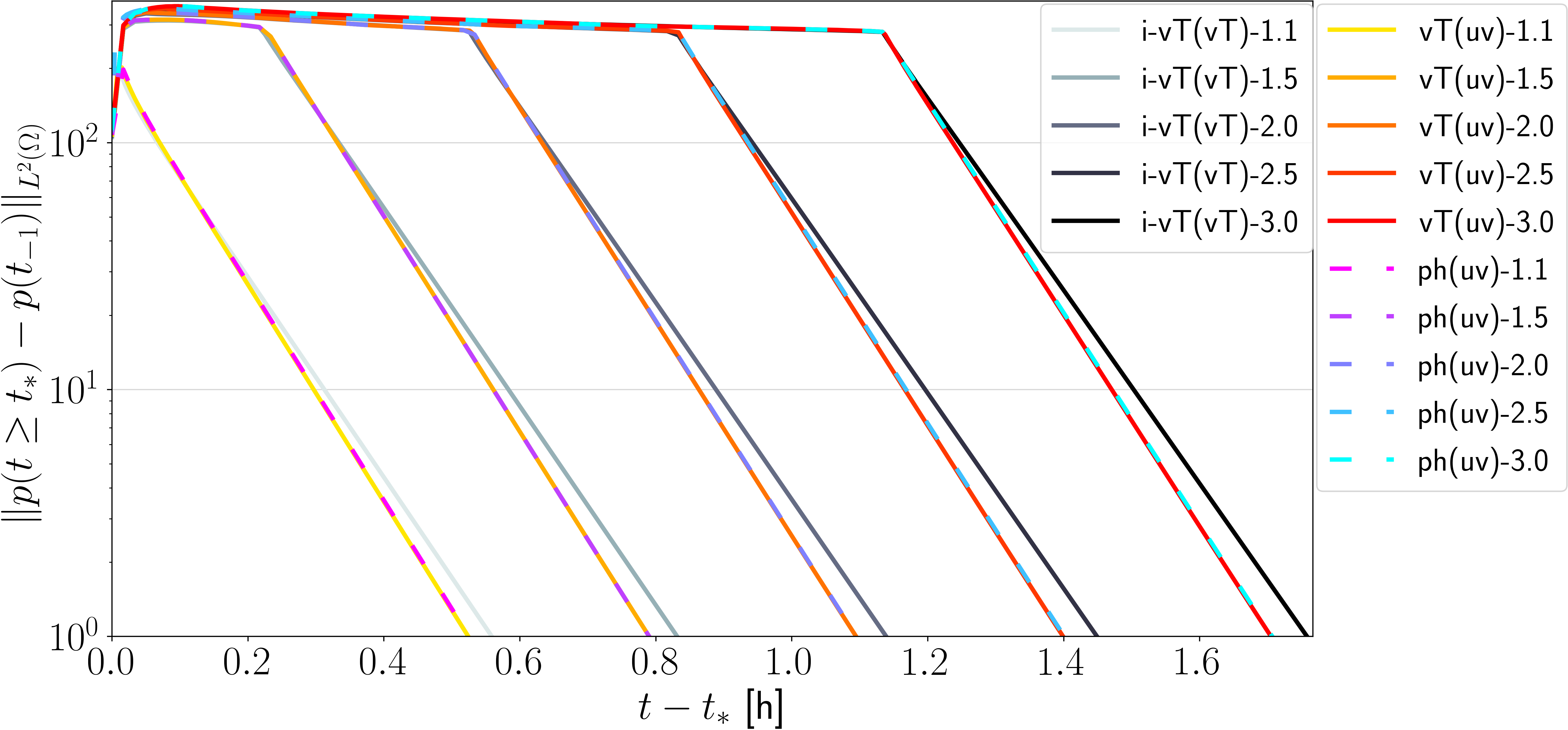}
    \caption{$L_2$-difference in pressure compared to the steady state solution before the fracture opening.}
    \label{fig:p-l2-over-tau}
\end{figure}

The $L^2$-norm reveals that the initial pressure disturbance extends over an increasingly large region surrounding the fracture as the aperture scaling factor increases.
Consequently, a larger portion of the computational domain contributes to the global pressure difference.
The temporal evolution of the $L^2$-difference relative to the steady-state solution preceding fracture opening is shown in \Cref{fig:p-l2-over-tau}.
The pressure transient is considered complete once the $L^2$-difference falls below unity.

A notable observation is that the $L^2$-difference initially increases, remains nearly constant for a period of time, and ultimately decays.
Its maximum is reached approximately when the decompression front has propagated through the region surrounding the injector, roughly two minutes after the fracture opening.
During this period, fluid is continuously injected into the reservoir, gradually replenishing the increased pore volume.
While gas remains present, the pressure recovers only slowly because much of the injected mass is accommodated by fluid compression and vapor condensation rather than by an increase in pressure.
Once the gas phase disappears, continued injection produces a more rapid pressure recovery until the system eventually reaches its new steady state.

In all cases, the pressure response is substantial.
Within the fracture and its immediate surroundings, the pressure decreases from approximately 10 \unit{\mega\pascal} to 0.8 \unit{\mega\pascal}.
Although the present model does not include a fully coupled momentum balance for the rock, these results suggest that mechanically induced phase change may strongly influence the subsequent mechanical response of the porous medium.
In particular, the pronounced pressure reduction may affect fracture closure, stress redistribution, and the propagation of elastic waves.
Quantifying these effects, however, requires a fully coupled flow–geomechanics formulation and is therefore left for future work.

Overall, all investigated quantities exhibit a monotonic dependence on the aperture scaling factor.
Larger fracture openings therefore lead to progressively stronger thermodynamic responses, while the comparison between the different formulations confirms that the persistent-variable framework yields consistent physical solutions independent of the chosen global equilibrium specification.
For this reason, simulations using the $\specIntEnergy\specVolume$-formulation as the global transport model have been omitted.
Since the thermal $\specVolume T$- and $p\specEnthalpy$-based formulations already recover the same physical solution, we do not expect the $\specIntEnergy\specVolume$-formulation to alter the physical results.
An additional comparison with the $\specIntEnergy\specVolume$-formulation would primarily confirm the same coordinate invariance while substantially increasing the number of simulation cases.
It would also introduce a considerably more challenging nonlinear system while providing little additional insight into the underlying physical processes considered in this study.

\WFfill
\WFclear
\FloatBarrier
\section{Conclusion}
We presented a persistent-variable formulation of thermal compositional flow based on independent specific-volume variables and isochoric local equilibrium calculations.
Starting from a fully coupled formulation of flow, transport, and thermodynamic equilibrium, we derived a hierarchy of volume-based models and demonstrated that classical pressure-based formulations arise naturally as reduced models through the elimination of the local thermodynamic degrees of freedom.
The resulting framework therefore establishes a direct mathematical connection between conventional pressure formulations and volume-balance approaches while maintaining thermodynamic consistency throughout.

The proposed formulation was motivated by applications in which rapid changes in pore volume influence the thermodynamic state of the fluid.
Rather than introducing volume as an auxiliary quantity, the present work promotes it to a transported thermodynamic state variable through which mechanical activity and fluid phase behavior can be coupled in a physically consistent manner.
This enables the resolution of transient phase changes induced solely by pore-volume alterations.
No additional constitutive assumptions relating pressure and pore volume are required.

To support such simulations, we introduced a physics-based nonlinear preconditioning strategy that models the instantaneous free expansion of the fluid following abrupt pore-volume changes.
Rather than representing an ad hoc numerical initialization, the preconditioner exploits the separation of physical time scales by resolving the thermodynamic transition before solving the coupled transport problem.
The numerical examples show that this strategy is capable of detecting and resolving transient mechanically induced phase change that is otherwise missed by conventional formulations or adaptive time stepping alone, while simultaneously improving the robustness of the nonlinear solver.

The thermal simulations demonstrate that mechanically induced phase transitions cannot, in general, be regarded as purely hydraulic phenomena.
Following rapid pore-volume expansion, the fluid undergoes a free expansion that initially cools the fluid and generates temperature gradients between the fracture and surrounding rock.
Although conductive heat transfer only weakly influences the overall duration of the transient, it alters the thermodynamic path followed by the fluid and leads to a different post-transient equilibrium state than predicted by an isothermal model.
These observations highlight the importance of consistently accounting for energy transport when mechanically induced phase transitions are considered.

Furthermore, the numerical experiments demonstrate that the equilibrium specification primarily determines the numerical properties of the nonlinear problem rather than the recovered physical solution.
While pressure- and volume-based formulations exhibit different nonlinear behavior and robustness, they converge to the same physical solution when solved accurately.
Moreover, the proposed nonlinear preconditioning strategy is not restricted to the presented volume-based formulation but can equally be combined with conventional pressure-based transport formulations.
This separation between the physical model, the equilibrium specification, and the nonlinear solution strategy constitutes one of the principal conceptual results of the present work.

The numerical results are consistent with previous experimental and modeling evidence that rapid fault dilation can induce substantial depressurization, partial vaporization, and delayed recovery through fluid recharge.
The present work extends these observations by embedding the rapid thermodynamic transition within a persistent-variable flow formulation, rather than prescribing it as a separate post-rupture initial condition.

Beyond the particular volume-based formulation considered here, the present work further supports the viewpoint that extensive thermodynamic state variables provide a natural and physically meaningful description of transport problems.
Unlike pressure and temperature, which are determined through constitutive relations, extensive quantities are directly governed by conservation laws, making them particularly well suited as primary variables in coupled multiphysics formulations.

Although our work considers prescribed fracture opening as a proof of concept, the formulation naturally lends itself to coupled flow–geomechanics problems.
By introducing independent extensive thermodynamic variables into the transport formulation, mechanically induced changes of the pore space can be propagated directly to the local equilibrium problem through a physically meaningful coupling quantity.
Extending the framework to account for fracture closure, contact mechanics, and slip requires coupling to momentum balance equations and appropriate constitutive models but does not alter the underlying thermodynamic formulation.
More broadly, the introduction of independent extensive state variables provides a general framework in which transport, thermodynamics, and mechanical deformation can be coupled while remaining compatible with established pressure-based formulations and numerical solution strategies.
We therefore view the presented formulation as a step toward a general persistent-variable framework for coupled subsurface multiphysics.
\section*{CRediT authorship contribution statement}

\noindent\textbf{Veljko Lipovac:} Conceptualization, Methodology, Software, Writing - Original Draft, Visualization.
\\
\textbf{Eirik Keilegavlen:} Conceptualization, Supervision, Writing - Review \& Editing.
\\
\textbf{Inga Berre:} Conceptualization, Supervision, Writing - Review \& Editing, Project administration, Funding acquisition.

\section*{Declaration of Competing Interest}
\noindent The authors declare that they have no known competing financial interests or personal relationships that could have appeared to influence the work reported in this paper.

\section*{Declaration of Generative AI Usage}
\noindent Generative artificial intelligence (GAI) tools were used for limited support during the preparation of this article.
The model OpenAI - GPT 5.5 was used for proofreading and generating editing suggestions.
The assistance was restricted to improving readability, phrasing, and grammar. 
Anthropic - Claude Opus 4.8 was used as a reading aid for individual literature sources and to assist in the preparation of bibliographic entries for \LaTeX.
All suggestions created by the GAI tools were carefully reviewed, verified, and edited to uphold scientific standards.
The author assumes full responsibility for the integrity, accuracy, and originality of the presented work.

\section*{Acknowledgments}
\noindent This result is part of a project has received funding from the European Research Council (ERC) under the European Union’s Horizon 2020 research and innovation program (grant agreement No 101002507).

\section*{Data availability}
\noindent The data and source code for the results presented herein is available, and the plots can be reproduced using a Docker container available at:\newline
\url{https://doi.org/10.5281/zenodo.21433114}.
\FloatBarrier
\newpage
\printbibliography

\end{document}